\RequirePackage[mathlines]{lineno} 
\documentclass[a4paper,11pt]{article}

\usepackage{jheppub} 
\usepackage{overpic}
\usepackage{amssymb}
\usepackage{amsmath}
\usepackage[T1]{fontenc} 
\usepackage{amsmath}
\usepackage{graphicx}
\usepackage{subfigure}
\usepackage{epstopdf}
\usepackage{rotating}

\usepackage{threeparttable}
\usepackage{multirow}
\usepackage{subfigure}
\usepackage{float}
\usepackage{verbatim}

\let\oldequation\equation
\let\oldendequation\endequation
\renewenvironment{equation}
  {\linenomathNonumbers\oldequation}
  {\oldendequation\endlinenomath}

\begin{document}

\title{\boldmath Analysis of the dynamics of the decay $D^{+}\to K_{S}^{0} \pi^{0} e^{+}\nu_{e}$}

\collaborationImg{\includegraphics[height=30mm,angle=90]{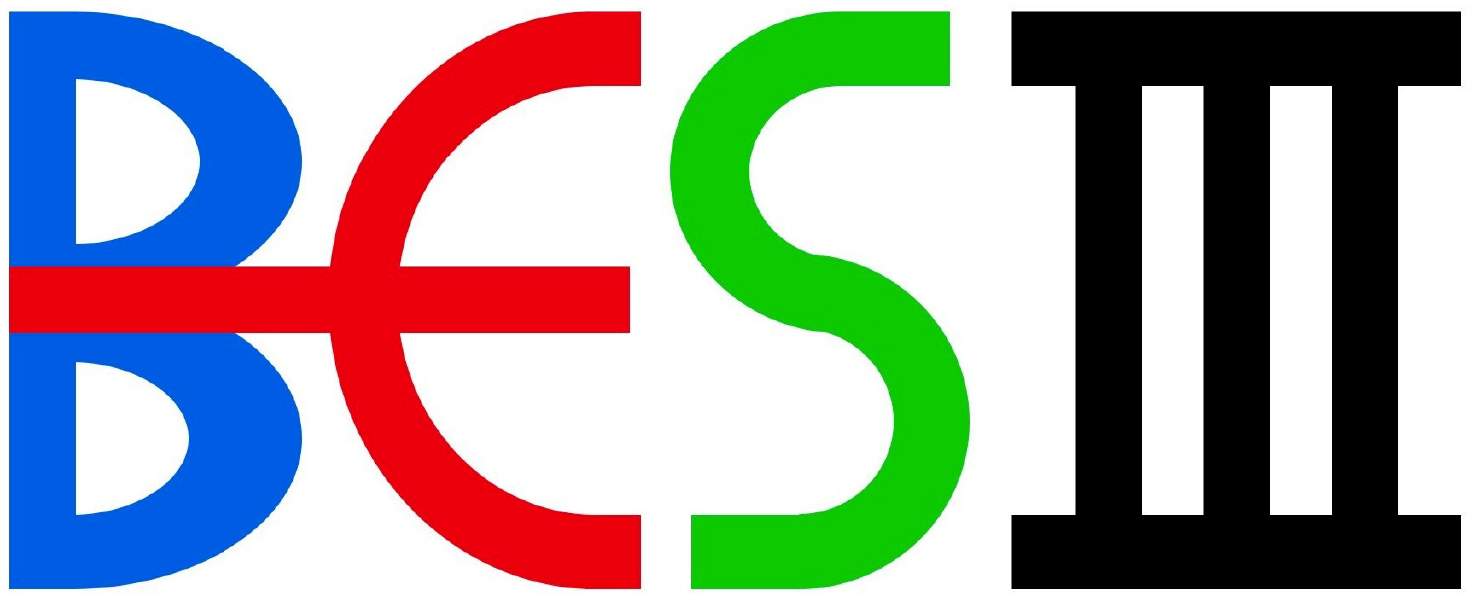}}
\collaboration{The BESIII collaboration}

\abstract{The branching fraction of $D^+\to K_{S}^{0} \pi^{0}e^+\nu_e$ is measured for the first time using $7.93~\mathrm{fb}^{-1}$ of $e^+e^-$ annihilation data collected at the center-of-mass energy $\sqrt{s}=3.773$~GeV with the BESIII detector operating at the BEPCII collider,  and is determined to be ${\mathcal B}$($D^+\to K_S^0\pi^0e^+\nu_e$) = $(0.881~\pm~0.017_{\rm stat.}~\pm~0.016_{\rm syst.})$\%.
Based on an analysis of the $D^+\to K_S^0\pi^0e^+\nu_e$ decay dynamics, we observe the $S\text{-}{\rm wave}$ and $P$-wave components with fractions of $f_{S\text{-}{\rm wave}}$ = $(6.13~\pm~0.27_{\rm stat.}~\pm ~0.30_{\rm syst.})\%$ and $f_{\bar K^{*}(892)^0}$ = $(93.88~\pm~0.27_{\rm stat.}~\pm~0.29_{\rm syst.})$\%, respectively.  From these results, we obtain the branching fractions ${\mathcal B}$($D^+\to (K_S^0\pi^0)_{S\text{-}{\rm wave}}~e^+\nu_e$)  = $(5.41~\pm~0.35_{\rm stat.}~\pm~0.37_{\rm syst.})\times10^{-4}$ and ${\mathcal B}$($D^+\to \bar K^{*}(892)^0e^+\nu_e$) = $(4.97~\pm~0.11_{\rm stat.}~\pm~0.12_{\rm syst.})$\%.
In addition, the hadronic form-factor ratios of $D^{+} \to \bar {K}^{*}(892)^0e^+\nu_e$ at $q^2=0$, assuming a single-pole dominance parameterization, are determined to be $r_V=\frac{V(0)}{A_1(0)}= 1.43~\pm~0.07_{\rm stat.}~\pm~0.03_{\rm syst.}$ and $r_2=\frac{A_2(0)}{A_1(0)}=0.72~\pm~0.06_{\rm stat.}~\pm~0.02_{\rm syst.}$.}

\maketitle
\flushbottom

\section{INTRODUCTION}
\label{sec:introduction}
\hspace{1.5em}
Semileptonic $D$ decays play an important role in our understanding of strong and weak effects in the charm sector~\cite{one,hard_weak1,hard_weak2}. 
Measurements of these semileptonic decay rates offer an opportunity to determine the Cabibbo-Kobayashi-Maskawa (CKM) matrix elements, which describe the quark-flavor mixing of the weak interaction in the standard model~(SM)~\cite{Cabibbo:1963yz, Kobayashi:1973fv,Ke:2023qzc}.
The strong interaction is characterized by the hadronic form factors that define the initial and final hadrons.
Various theoretical models, such as Lattice QCD~\cite{Lubicz:1992,Bernard:1991bz,Abada:1994,Bowler:1994zr,Bhattacharya:1994db,APE:1994kxx}, the quark model~\cite{Wirbel:1985ji,Isgur:1988gb,Gilman:1989uy}, and QCD sum rules~\cite{theo:Sum}, make predictions for the branching fractions~(BFs) or hadronic form factors of $D\to K^{*}$ transitions. The measurements of the BFs or hadronic form factors of $D^+\to \bar K^{*0}e^+\nu_e$ are therefore valuable for testing and validating these theoretical calculations~\cite{Lubicz:1992,Bernard:1991bz,Abada:1994,Bowler:1994zr,Bhattacharya:1994db,APE:1994kxx,Wirbel:1985ji,Isgur:1988gb,Gilman:1989uy,theo:Sum}.

\begin{figure}[htbp]
\centering
\includegraphics[width=0.5\linewidth]{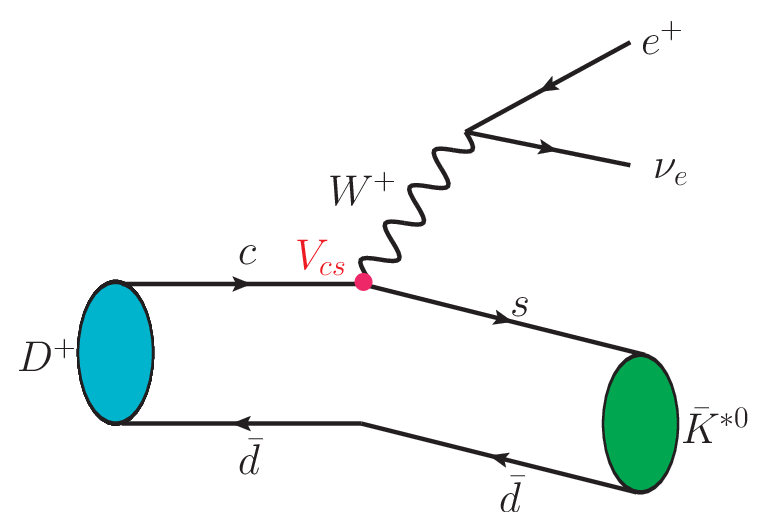}
\caption{\footnotesize The Feynman diagram of the $D^+\to \bar K^{*0}e^{+}\nu_{e}$ decay.}
\label{fig:fey}
\end{figure}

Experimental measurements of the BFs and form factors of $D^+\to \bar K^{*0}e^{+}\nu_{e}$ via $\bar K^{*0}\to K^-\pi^+$ have been reported by various experiments, including MARKIII~\cite{MARK-III:1990bbt}, CLEO-c~\cite{four}, BaBar~\cite{five}, and BESIII~\cite{six1}, and are summarized by the Particle Data Group (PDG)~\cite{ref::pdg2022}. However, no experimental study of $D^+\to \bar K^{*0}e^{+}\nu_{e}$ via $\bar K^{*0}\to \bar K^0\pi^0$ has yet been performed. The $D^+\to K_S^0\pi^0e^+\nu_e$ decay is expected to be dominated by $D^+\to \bar K^{*0}e^+\nu_e$, which is mediated via the Feynman diagram shown in Fig.~\ref{fig:fey}. 
In this paper, we present the first study  of the $D^+\to K_{S}^{0}\pi^{0}e^{+}\nu_e$ decay dynamics. This analysis is based on a data sample of $e^+e^-$ collisions collected at a center-of-mass energy of $\sqrt s =$ 3.773 GeV  by the BESIII detector operating at the BEPCII collider in 2010, 2011, and 2021, and correspond to a total integrated luminosity of 7.93 fb$^{-1}$~\cite{Ablikim:2013ntc}.
Throughout this paper, the charge conjugate channels are always implied.

\section{BESIII DETECTOR AND MONTE CARLO SIMULATION}
\label{sec:detector}
\hspace{1.5em}

The BESIII detector~\cite{Ablikim:2009aa} records symmetric $e^+e^-$ collisions
provided by the BEPCII storage ring~\cite{Yu:IPAC2016-TUYA01}
in the center-of-mass energy range from 2.00 to 4.95~GeV,
with a peak luminosity of $1.1 \times 10^{33}\;\text{cm}^{-2}\text{s}^{-1}$
achieved at $\sqrt{s} = 3.77\;\text{GeV}$.
BESIII has collected large data samples in this energy region~\cite{Ablikim:2019hff,EcmsMea,EventFilter}. The cylindrical core of the BESIII detector~\cite{detvis} covers 93\% of the full solid angle and consists of a helium-based
 multilayer drift chamber~(MDC), a plastic scintillator time-of-flight
system~(TOF), and a CsI(Tl) electromagnetic calorimeter~(EMC),
which are all enclosed in a superconducting solenoidal magnet
providing a 1.0~T magnetic field.

The solenoid is supported by an
octagonal flux-return yoke with resistive plate counter muon
identification modules interleaved with steel.
The charged-particle momentum resolution at $1~{\rm GeV}/c$ is
$0.5\%$, and the
d$E$/d$x$
resolution is $6\%$ for electrons
from Bhabha scattering. The EMC measures photon energies with a
resolution of $2.5\%$ ($5\%$) at $1$~GeV in the barrel (end-cap)
region. The time resolution in the TOF barrel region is 68~ps, while
that in the end-cap region of data collected before 2015 is 110~ps.
The end-cap TOF system was upgraded in 2015 using multigap resistive plate chamber
technology, providing a time resolution of 60~ps, which benefits 83\% of the data used in this analysis~\cite{etof}.

Simulated event samples produced with the {\sc geant4}-based~\cite{geant4} Monte Carlo (MC) package, which
includes the geometric description of the BESIII detector and the
detector response, are used to determine the detection efficiency
and estimate backgrounds. The simulation includes the beam-energy spread and initial-state radiation in the $e^+e^-$
annihilations modeled with the {\sc kkmc} generator~\cite{kkmc}.
An inclusive MC sample includes the production of $D\bar{D}$
pairs (including quantum coherence for the neutral $D$ channels), the non-$D\bar{D}$ decays of the $\psi(3770)$, the initial-state radiation
production of the $J/\psi$ and $\psi(3686)$ states, and the
continuum processes incorporated in {\sc kkmc}~\cite{kkmc}.
All particle decays are modelled with {\sc
evtgen}~\cite{evtgen} using the BFs either taken from the
Particle Data Group~\cite{ref::pdg2022}, when available, or otherwise estimated with {\sc lundcharm}~\cite{lundcharm}. Final-state radiation from charged particles is incorporated with the {\sc
photos} package~\cite{photos}.

In this paper, the inclusive MC sample is used to determine the 
selection efficiencies of the tag side and to estimate the backgrounds. The signal MC samples of $D^+ \to K_S^0\pi^0e^+\nu_e$, which are used to determine the signal efficiency, are simulated with the parameters obtained from the amplitude-analysis fit.

\section{METHOD}
\hspace{1.5em}
At $\sqrt s=3.773$\rm \,GeV, $D^+D^-$ meson pairs are produced from $\psi(3770)$ decays without accompanying hadrons, providing an ideal opportunity to study semileptonic $D^+$ decays using the double-tag~(DT) method~\cite{mark3}.
Initially, single-tag~(ST) $D^-$ mesons are reconstructed via the decays $D^-\to K^{+}\pi^{-}\pi^{-}$, $K^0_{S}\pi^{-}$, $K^{+}\pi^{-}\pi^{-}\pi^{0}$, $K^0_{S}\pi^{-}\pi^{0}$, $K^0_{S}\pi^{+}\pi^{-}\pi^{-}$, and $K^{+}K^{-}\pi^{-}$.
Subsequently, the semileptonic $D^+$ candidates are reconstructed by using the remaining tracks that were not utilized in the ST selection.
An event in which the semileptonic decay $D^+\to K_{S}^{0}\pi^{0}e^{+}\nu_{e}$ is reconstructed in the system recoiling against the ST $D^-$ mesons is referred to as a DT event.
The BF of $D^+\to K_{S}^{0}\pi^{0}e^{+}\nu_{e}$ is determined by
\begin{equation}
    \label{br}
   {
{\mathcal B}_{\rm SL} = \frac{N_{\rm DT}}{N^{\rm tot}_{\rm ST} \bar \epsilon_{\rm sig} {\mathcal B}_{K_S^{0}}{\mathcal B}_{\pi^{0}}},}
    \end{equation}
where $N_{\rm ST}^{\rm tot}$ and $N_{\rm DT}$ represent the yields of the ST $D^-$ mesons and the DT signal events in data, respectively; $\mathcal B_{K^0_S}$ and $\mathcal B_{\pi^0}$ are the BFs of $K^0_S\to \pi^+\pi^-$ and $\pi^0\to \gamma \gamma$, respectively, as reported by
the PDG~\cite{ref::pdg2022}. $\bar \epsilon_{\rm sig}$ denotes the average signal efficiency weighted by the measured yield of tag mode $i$ in the data, i.e.,
   \begin{equation}
    \label{eff}
   {
\bar{{\mathcal \epsilon}}_{\rm sig} = \frac{\sum_i (N^i_{\rm ST}\cdot \epsilon^i_{\rm DT}/\epsilon^i_{\rm ST})}{N^{\rm tot}_{\rm ST}},}
    \end{equation}
where $N^i_{\rm ST}$ is the yield of the observed ST candidates in data, $\epsilon^i_{\rm ST}$ is the efficiency of reconstructing the ST mode $i$ (called the ST efficiency),
and $\epsilon^i_{ \rm DT}$ is the efficiency of finding the ST mode $i$ and the $D^+\to K_{S}^{0}\pi^{0}e^{+}\nu_{e}$ decay simultaneously (called the DT efficiency).

\section{SINGLE-TAG SELECTION}
\hspace{1.5em}
For each charged track (except those used for $K^0_S$ reconstruction), the polar angle with respect to the MDC axis ($\theta$) must satisfy $|\!\cos\theta|<0.93$, and the point of closest approach to the interaction point~(IP) must lie within 1\,cm in the plane perpendicular to the MDC axis and within 10\,cm along the MDC axis.
Charged tracks are identified using the d$E$/d$x$ and TOF information, from which the combined confidence levels under the pion and kaon hypotheses are separately computed. The charged tracks are then assigned to the particle type with the higher probability.

Candidates for $K_S^0$ are formed from pairs of oppositely charged tracks. For these two tracks, the distance of closest approach to the IP is required to be less than 20\,cm along the MDC axis. There are no limitations on the distance of closest approach in the transverse plane or particle-identification (PID) criteria for these tracks. The two charged tracks are constrained to originate from a common vertex, which must be at least twice the vertex resolution away from the IP in terms of flight distance. The quality of the vertex fits (primary vertex fit and second vertex fit) is ensured by a requirement on the $\chi^2$ ($\chi^2 < 100$).
The invariant mass of the $\pi^+\pi^-$ pair is required to be within $(0.487,0.511)$~GeV/$c^2$.

Neutral-pion candidates are reconstructed via the $\pi^0\to\gamma\gamma$ decay. The EMC time deviation from the event start time is required to lie within [0,\,700]\,ns.  The energy deposited in the EMC  is required to be greater than 25 MeV in the barrel region ($|\!\cos\theta|<0.80$) and 50~MeV in the end cap ($0.86<|\!\cos\theta|<0.92$).
The opening angle between the photon candidate and the nearest charged track in the EMC is required to be greater than $10^{\circ}$. At least one of the photons is required to be detected
in the barrel EMC as the end caps have significantly worse resolution than the barrel. For each $\pi^0$ candidate, the invariant mass of the photon pair is required to be within $(0.115,\,0.150)$\,GeV$/c^{2}$. To improve the momentum resolution, a kinematic fit is performed in which the $\gamma\gamma $ invariant mass is constrained to the known $\pi^{0}$ mass~\cite{ref::pdg2022}, and the $\chi^{2}$ of the fit is required to be less than 50. The four-momentum of the $\pi^0$ candidate returned by this kinematic fit is used in the subsequent analyses.

To distinguish the ST $D^-$ mesons from combinatorial backgrounds, we define the energy difference $\Delta E\equiv E_{D^-}-E_{\mathrm{beam}}$ and the beam-constrained mass $M_{\rm BC}\equiv\sqrt{E_{\mathrm{beam}}^{2}/c^{4}-|\vec{p}_{D^-}|^{2}/c^{2}}$, where $E_{\mathrm{beam}}$ is the beam energy, and $E_{D^-}$ and $\vec{p}_{D^-}$ are the total energy and momentum of the ST $D^-$ meson in the $e^+e^-$ center-of-mass frame.
If there is more than one $D^-$ candidate for each ST mode, the one with the least $|\Delta E|$ is retained for the subsequent analysis.
The $\Delta E$ requirements and ST efficiencies are summarized in Table~\ref{ST:realdata}.

\begin{figure}[htbp]\centering
	\setlength{\abovecaptionskip}{-1pt}
	\setlength{\belowcaptionskip}{10pt}
\includegraphics[width=12.0cm]{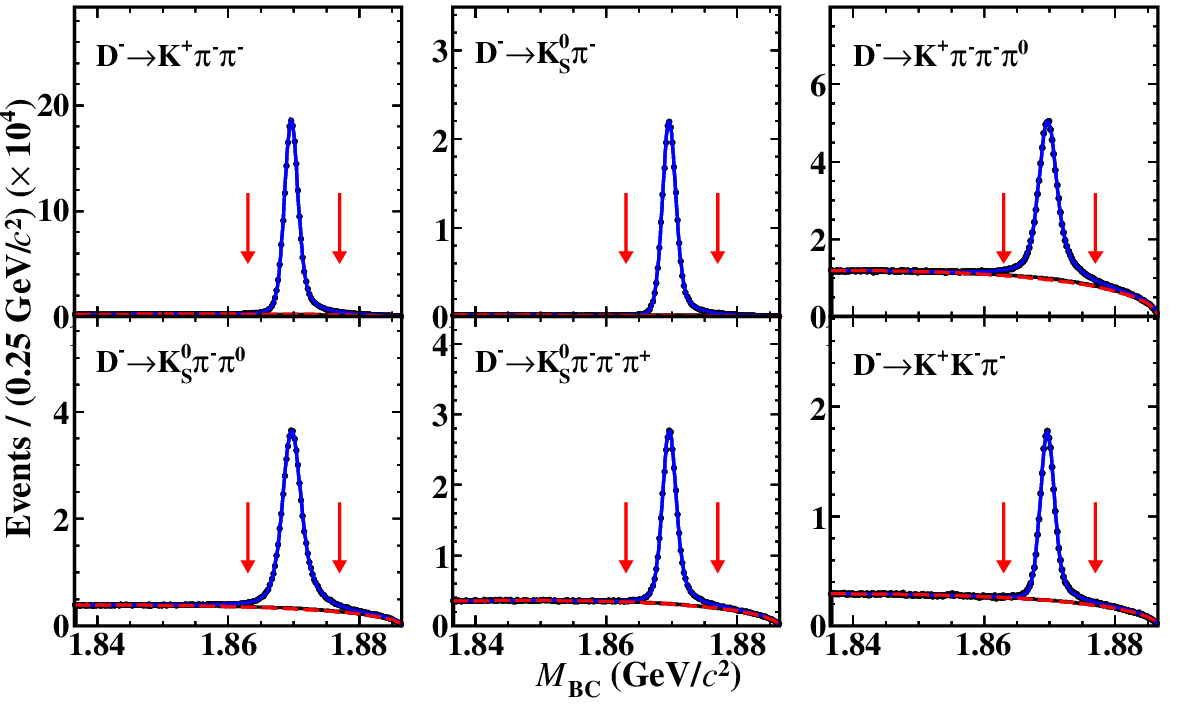}
\caption{
Fits to the $M_{\rm BC}$ distributions of the ST $D^-$ candidates.
In each plot, the points with error bars are data, the blue curves are the best fits, and the red dashed curves are the fitted combinatorial-background shapes. The pairs of red arrows show the $M_{\rm BC}$ signal windows.}\label{fig:datafit_Massbc}
\end{figure}

For each tag mode, the yield of ST $D^-$ mesons is obtained by fitting the corresponding $M_{\rm BC}$ distribution. In the fit, the signal shape is described by MC-simulated signal shape convolved with a double-Gaussian function.
The background shape is modeled by the ARGUS function~\cite{argus}, with the endpoint fixed at 1.8865~GeV/$c^{2}$ corresponding to $E_{\rm beam}$.
Figure~\ref{fig:datafit_Massbc} shows the results of the fits to the $M_{\rm BC}$ distributions of the accepted ST candidates in data for different tag modes. The candidates with $M_{\rm BC}$ lying within $(1.863,1.877)$ GeV/$c^2$ for $D^-$ tags are kept for the subsequent analysis. When summing over the tag modes  the total yield of ST $D^-$ mesons is determined to be $(4149.9\pm2.3_{\rm stat.})\times 10^3$.

\begin{table}
\renewcommand{\arraystretch}{1.2}
\centering
\caption {The $\Delta E$ requirements, the measured ST $D^-$ yields in the data~($N_{\rm ST}^{i}$), the ST efficiencies~($\epsilon_{\rm ST}^{i}$) and DT efficiencies~($\epsilon_{\rm DT}^{i}$). The uncertainties are statistical only.}
\begin{tabular}{ccccc}
\hline
\hline
Tag mode & $\Delta E~$(GeV)  &  $N_{\rm ST}^{i}~(\times 10^3)$  &  $\epsilon_{\rm ST}^{i}~(\%)$&$\epsilon_{\rm DT}^{i}~(\%)$       \\\hline
$D^-\to K^+\pi^-\pi^-$                   &  $(-0.025,0.024)$ & $2164.0\pm1.5$&$51.17\pm0.01$&$8.55\pm0.01$\\
$D^-\to K^{0}_{S}\pi^{-}$                &  $(-0.025,0.026)$ & $250.4\pm0.5$&$50.74\pm0.02$&$8.45\pm0.03$\\
$D^-\to K^{+}\pi^{-}\pi^{-}\pi^{0}$      &  $(-0.057,0.046)$ & $689.0\pm1.1$&$25.50\pm0.01$&$3.71\pm0.01$\\
$D^-\to K^{0}_{S}\pi^{-}\pi^{0}$         &  $(-0.062,0.049)$ & $558.4\pm0.9$&$26.28\pm0.01$&$3.92\pm0.01$\\
$D^-\to K^{0}_{S}\pi^{-}\pi^{-}\pi^{+}$  &  $(-0.028,0.027)$ & $300.5\pm0.6$&$29.01\pm0.01$&$4.23\pm0.02$\\
$D^-\to K^{+}K^{-}\pi^{-}$               &  $(-0.024,0.023)$ & $187.3\pm0.5$&$41.06\pm0.02$&$6.89\pm0.03$\\
\hline
\hline
          \end{tabular}
          \label{ST:realdata}
          \end{table}

\vspace{-0.0cm}
\vspace{-0.0cm}

\section{DOUBLE-TAG SELECTION}
\hspace{1.5em}
The candidates for $D^+\to K_{S}^{0}\pi^{0}e^{+}\nu_{e}$ decays are selected from the remaining tracks in the presence of the tagged $D^-$ candidates.
It is required that there are no extra good charged tracks~($N^{\rm charge}_{\rm extra}$) in addition to those used to construct the  $D^+\to K_{S}^{0}\pi^{0}e^{+}\nu_{e}$ candidate.

Candidates for $K_{S}^{0}$ and $\pi^0$ are selected with the same criteria as those used in the tag selection. The positron is identified using the TOF, d$E$/d$x$, and EMC measurements to calculate the combined confidence levels $CL_{e}$, $CL_{K}$, and $CL_{\pi}$ for electron, kaon and pion hypotheses, respectively. The positron candidate is required to satisfy $CL_{e} > 0.8\times (CL_{e}+CL_{\pi}+CL_{K})$, $CL_{e} >0.001$, and $E_{\rm EMC}/p_{\rm MDC}>0.8c$. Here, $E_{\rm EMC}$ is the energy deposited in the EMC, while $p_{\rm MDC}$ is the momentum measured by the MDC.

To reject the backgrounds from hadronic decays involving a $\pi^0$, such as $D^+ \to K_S^0 \pi^0 \pi^+ \pi^0$, the maximum energy of any extra photons($E_{\text{extra~}\gamma}^{\rm max}$) which has not been used in the event selection is required to be less than 0.25~GeV for the $D^+\to K_{S}^{0}\pi^{0}e^{+}\nu_{e}$ selection.
To suppress possible contamination from the hadronic decays $D^+\to K_S^0\pi^0\pi^+$, where
the pion is misidentified as a positron, the invariant mass of $K_S^{0}\pi^{0}e^{+}$~($M_{K_S^{0}\pi^{0}e^{+}}$) is required to be less than 1.76 GeV/$c^2$.
To suppress background events $D^+\to K_{S}^{0}e^{+}\nu_{e}$ due to an additional fake $\pi^{0}$, a new variable $U_{\rm miss}^{\rm K_S^0e^+\nu_e}$ is used. This variable is defined as $U_{\rm miss}^{\rm K_S^0e^+\nu_e}\equiv E_{\rm miss}^{\rm K_S^0e^+\nu_e}-p_{\rm miss}^{\rm K_S^0e^+\nu_e}c$, where $E_{\rm miss}^{\rm K_S^0e^+\nu_e}=E_{\rm beam}-E_{K_{S}^{0}}-E_{e^{+}}$ and $p_{\rm miss}^{\rm K_S^0e^+\nu_e}=|\vec{p}_{D}-{\vec p}_{K_{S}^{0}}-{\vec p}_{e^{+}}|$. Here, $E_{\rm miss}^{\rm K_S^0e^+\nu_e}$ and $p_{\rm miss}^{\rm K_S^0e^+\nu_e}$ represent the total energy and the momentum of all missing particles in the DT events, respectively. $U_{\rm miss}^{\rm K_S^0e^+\nu_e}$ is required to be larger than 0.04~GeV.
These requirements have been optimized using a Figure of Merit defined as $\frac{S}{\sqrt{S+B}}$. Here, $S$ and $B$ denote the signal and background yields from the normalized inclusive MC sample.

The neutrino cannot be directly detected by the BESIII detector.
To select semileptonic signal candidates, we define $U_{\mathrm{miss}}\equiv E_{\mathrm{miss}}-|\vec{p}_{\mathrm{miss}}|c$, where $E_{\mathrm{miss}}$ and $\vec{p}_{\mathrm{miss}}$
are the missing energy and momentum, respectively, of the DT event in the $e^+e^-$ center-of-mass frame.
These quantities are calculated by $E_{\mathrm{miss}}\equiv E_{\mathrm{beam}}-E_{K_S^0}-E_{\pi^{0}}-E_{e^{+}}$ and $\vec{p}_{\mathrm{miss}}\equiv\vec{p}_{D^+}-\vec{p}_{K_S^0}-\vec{p}_{\pi^{0}}-\vec{p}_{e^{+}}$, where $E_{K_S^{0}\,(\pi^0)\,(e^+)}$ and $\vec{p}_{K_S^{0}\,(\pi^0)\,(e^+)}$ are the measured energies and momenta of the $K_S^{0}\,(\pi^0)\,(e^+)$ candidates, respectively, and $\vec{p}_{D^+}\equiv-\hat{p}_{D^-} \sqrt{E_{\mathrm{beam}}^{2}/c^{2}-m_{D^-}^{2} c^{2} }$, where
$\hat{p}_{D^-}$ is the unit vector in the momentum direction of the ST $D^-$ meson and $m_{D^-}$ is the nominal $D^-$ mass~\cite{ref::pdg2022}.
In the presence of the ST $D^-$ mesons, the average signal efficiency for $D^+\to K_{S}^{0}\pi^{0}e^{+}\nu_{e}$ is determined to be $(15.40\pm0.01)\%$, which is corrected  according to the discussions in Sec.~\ref{corr}.

\section{BRANCHING FRACTION}\label{bf}
\begin{figure}[htbp]\centering
	\setlength{\abovecaptionskip}{-1pt}
	\setlength{\belowcaptionskip}{10pt}
\includegraphics[width=8.0cm]{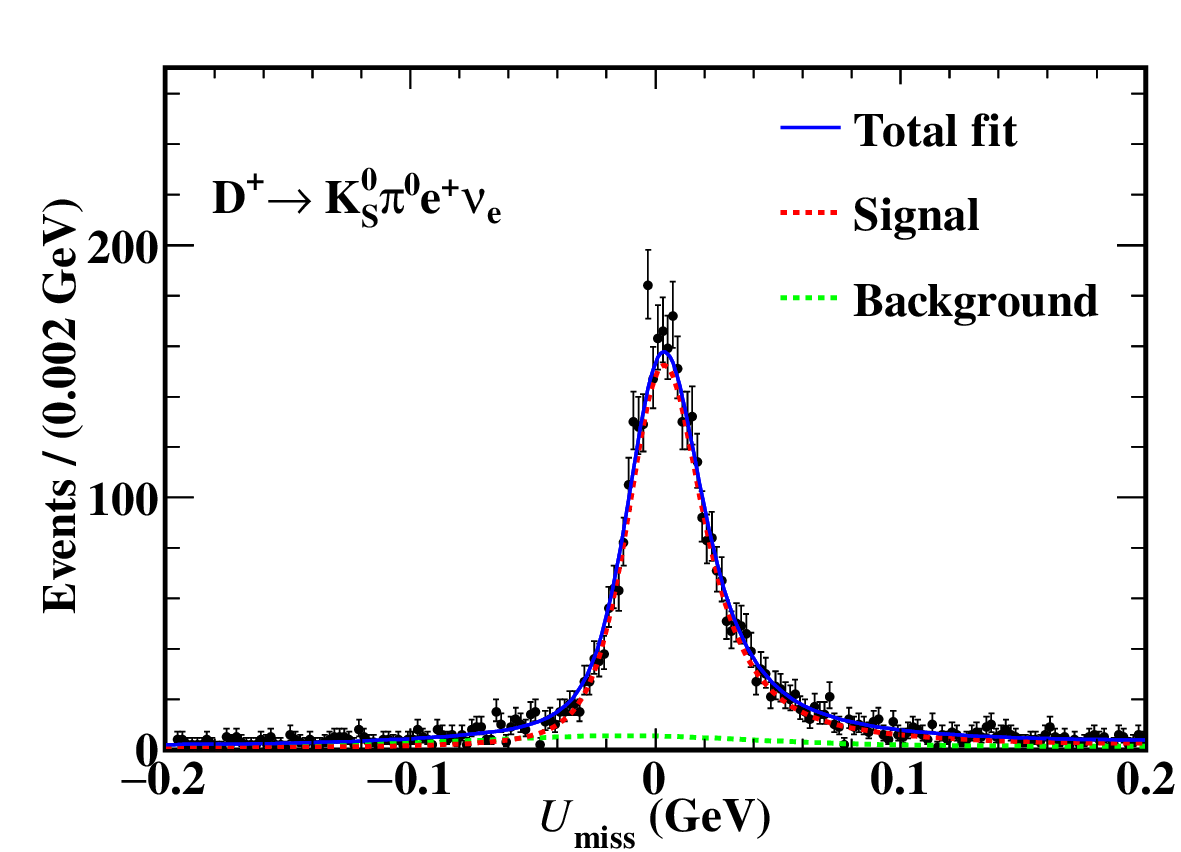}
\caption{ Fit to the $U_{\rm miss}$ distribution of the selected $D^+\to K_{S}^{0}\pi^{0}e^{+}\nu_{e}$ candidate events.  
}
\label{fig:Umissfit}
\end{figure}

\hspace{1.5em}
To extract the signal yield, an unbinned extended maximum-likelihood fit is performed to the $U_{\rm miss}$ distribution of the accepted candidates for $D^+\to K_{S}^{0}\pi^{0}e^{+}\nu_{e}$. In the fit, the signal shape is described by the MC-simulated shape convolved with a Gaussian function with floating parameters to account for the difference in resolution between data and MC simulation. The background shape is derived from the inclusive MC sample. The numbers of signal and background events are free parameters in the fit. The signal yields are determined to be $(3852\pm75)$ by fitting the $U_{\rm miss}$ distribution in data, and the fitting result is shown in Fig.~\ref{fig:Umissfit}. With Eq.~(\ref{br}), the BF of $D^+\to K_S^0\pi^0e^+\nu_e$ is determined to be ($0.881\pm0.017_{\rm stat.}\pm0.016_{\rm syst.}$)\%.
\vspace{-0.0cm}
\vspace{-0.0cm}

\section{SYSTEMATIC UNCERTAINTY OF BRANCHING FRACTION}\label{corr}
\hspace{1.5em}
The DT method has the benefit that many systematic uncertainties due to the ST selection cancel in the ratio used to perform the BF measurement. Those systematic uncertainties that do not cancel, and those associated with the DT selection, are discussed below.

\begin{itemize}
\item[\bf (a)] {$N_{\rm ST}^{\rm tot}$}: The uncertainty related to the ST yield $N_{\rm ST}^{\rm tot}$ is assigned to be 0.1\% by varying the signal and background shapes. The signal shape is varied by changing the truth-matched angle from $15^{\circ}$ to $10^{\circ}$ or $20^{\circ}$, while the background shape is varied by changing the endpoint of ARGUS  function from 1.8865 to 1.8863 or 1.8867 GeV/$c^2$.  Additionally, the parameters of one of the two Gaussian functions used in the signal description is left free in the fit. 
    Different fits are then performed of the $M_{\rm BC}$ distribution for each ST mode for both data and inclusive MC sample, and the difference in $N_{\rm ST}/\epsilon_{\rm ST}$ is assigned as the systematic uncertainty.

\item[\bf (b)] {$e^\pm$ tracking, PID and $E_{\rm EMC}/p_{\rm MDC}$}: The systematic uncertainties arising from the knowledge of the  tracking, PID efficiencies of $e^\pm$, and $E_{\rm EMC}/p_{\rm MDC}$ are assessed from a control sample of  radiative Bhabha scattering events. The difference of the selection efficiency between data and MC simulation, 0.5\%, is assigned as the corresponding systematic uncertainty.

\item[\bf (c)] {$K_{S}^0$ reconstruction}: The systematic uncertainty of the $K_S^0 \to \pi^+\pi^-$ reconstruction is considered to have two contributions. The $\pi^\pm$ tracking efficiencies are evaluated using  constrol samples of  $D^0\to K^-\pi^+$, $K^-\pi^+\pi^+\pi^-$ vs. $\bar D^0\to K^+\pi^-$, $K^+\pi^-\pi^-\pi^+$, ane $D^+ \to K^-\pi^+\pi^+$ vs. $D^- \to K^+\pi^-\pi^-$ events.  The event is fully reconstructed with the exception of one pion, the presence of which is inferred from the magnitude of the missing invariant mass in the event.  The tracking efficiency is then measured directly by checking whether this pion is reconstructed.
The difference in $\pi^\pm$ tracking efficiencies between data and MC simulation is determined. The efficiencies associated with the $K_S^0$ mass window, primary vertex and $K_S^0$ decay vertex fit are studied in  hadronic $D\bar D$ events, with $D^0$ decaying into $K_S^0\pi^+\pi^-$, $K_S^0\pi^+\pi^-\pi^0$ and $K_S^0\pi^0$, or $D^+$ decaying into $K_S^0\pi^+$, $K_S^0\pi^+\pi^0$, and $K_S^0\pi^+\pi^+\pi^-$. After correcting the MC efficiencies to the data, the residual statistical uncertainties of the data/MC differences are assigned as the systematic uncertainties.

\item[\bf (d)] {$\pi^0$ reconstruction}: The uncertainty from the $\pi^{0}$ reconstruction is assigned to be 1.0\% after studying a control sample of $D^0 \to K^- \pi^+\pi^0$ decays. The difference between data and MC simulation is $(99.4\pm1.0)\%$. After correcting the MC efficiencies to data, we assign 1.0\% as the uncertainty for $\pi^0$ reconstruction.

\item[\bf (e)] {Assumed $\mathcal B$}: The uncertainties in the assumed BFs of $K_{S}^{0}\to \pi^+\pi^-$ and $\pi^0\to \gamma\gamma$ are 0.07\% and 0.03\%~\cite{ref::pdg2022}, respectively.

\item[\bf (f)] {MC sample size}: The uncertainty due to the MC sample size is  0.1\%.

\item[\bf (g)] {$E^{\rm max}_{\rm extra\gamma}$ and $N^{\rm good}_{\rm charge}$ requirements}: The systematic uncertainty from the $E^{\rm max}_{\rm extra\gamma}$ and $N^{\rm good}_{\rm charge}$ requirements is estimated to be 0.8\%,
after performing DT samples of $D^+\to K_S^0e^+\nu_e$ decays  versus the tag modes, reconstructed as in the baseline analysis.

\item[\bf (h)] {$M_{K_{S}^{0}\pi^{0}e^+}$ requirement}: The uncertainty associated with the $M_{K_S^0\pi^0e^+}$ requirement is studied by varying its value by $\pm 10$ MeV/$c^2$, following the method defined in Refs.~\cite{BESIII:2017pez,BESIII:2018nzb,BESIII:2021uqr}. The maximum change in the signal efficiency is taken as the associated systematic uncertainty.

\item[\bf (i)] {$U_{\rm miss}^{K_{S}^{0}e^+\nu_e}$ requirement}: The efficiencies of the $U_{\rm miss}^{K_{S}^{0}e^+\nu_e}$ requirement are greater than 99\% and the difference of these efficiencies between data and MC simulation is negligible.

\item[\bf (j)] {MC model}: The signal MC samples in this study are generated with the generator developed with the parameters obtained in Section.~\ref{pwa}. New MC samples are generated in which the input form factors $r_V$ and $r_2$ are varied by $\pm 1\sigma$ of their statistical uncertainty around their baseline values. The largest change in the signal efficiency, 1.0\%, is taken as the systematic uncertainty in the MC generator.
\end{itemize}

The total systematic uncertainty is obtained by adding the individual components in quadrature ($\delta_{\rm syst}$), assuming that all sources are independent of each other.
Table~\ref{sys} summarizes the sources of the systematic uncertainties in the BF measurement.

\begin{table}[h!]
      \centering
       \caption{Relative systematic uncertainties~(in \%) in the BF measurement.}
\begin{tabular}{cc}
\hline
Source & Uncertainty  \\
\hline
$N_{\rm ST}^{\rm tot}$                         &0.1 \\
$e^\pm$ tracking                               &0.5 \\
$e^\pm$ PID and $E_{\rm EMC}/p_{\rm MDC}$      &0.5 \\
$K_{S}^0$ reconstruction                       &1.0 \\
$\pi^0$ reconstruction                         &1.0 \\
Assumed $\mathcal B$                            &0.1 \\
MC samples size                                 &0.1 \\
$E^{\rm max}_{\rm extra\gamma} \& N^{\rm good}_{\rm charge}$ requirements    &0.8 \\
$M_{K_{S}^{0}\pi^{0}e^+}$ requirement          &0.2 \\
$U_{\rm miss}^{K_{S}^{0}e^+\nu_e}$ requirement&negligible\\
MC model &1.0 \\
\hline
Total                                          &2.1\\
\hline  \hline
\end{tabular}
\label{sys}
\end{table}

\section{AMPLITUDE ANALYSIS OF $D^+\to K_{S}^{0}\pi^{0}e^{+}\nu_{e}$}\label{pwa}
\begin{figure*}[htbp]\centering
	\setlength{\abovecaptionskip}{-1pt}
	\setlength{\belowcaptionskip}{10pt}
\includegraphics[width=12cm]{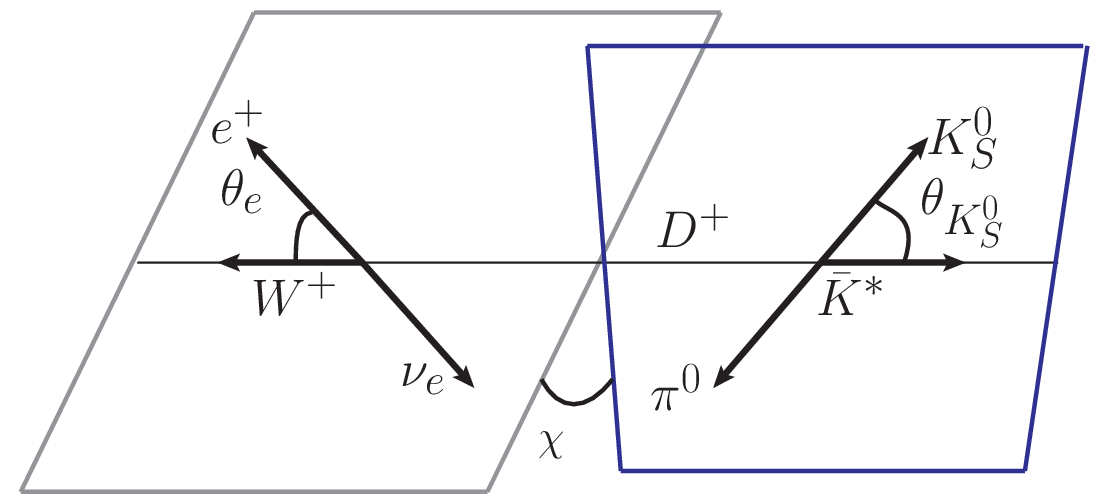}
\caption{Definition of the angular variables in the $D^+ \to K_S^0 \pi^0 e^+ \nu_e$ decay.
}
\label{fig:Kine}
\end{figure*}

\hspace{1.5em}
To prepare a sample for the study of the decay dynamics of $D^+\to K_{S}^{0}\pi^{0}e^{+}\nu_{e}$,  an additional requirement of $U_{\rm miss} \in [-0.04, 0.06]$~GeV is imposed on the events selected for the branching-fraction measurement.  With this additional requirement, we observe 3566 candidates in data with a background fraction of $(6.54\pm0.64)$\%. The four-body decay $D^+ \to K_S^0 \pi^0 e^+ \nu_e$ can be uniquely characterized by  five kinematic variables \cite{pwa-kinvariA, pwa-kinvariB}, as illustrated in Fig.~\ref{fig:Kine}: the invariant-mass squared of $K_S^0 \pi^0$~($m^{2}$), the invariant-mass squared of $e^+ \nu_e$~($q^{2}$),
the angle between the three-momentum of the $K_S^0$ in the $K_S^0\pi^0$ rest frame and the direction of flight of the $K_S^0\pi^0$ in the $D^+$ rest frame~($\theta_{K_S^0}$), the angle between the three-momentum of $e^+$ in the $e^+ \nu_e$ rest frame and the direction of flight of the $e^+ \nu_e$ in the $D^+$ rest frame ($\theta_{e}$), and the angle between the two decay planes~($\chi$).
The sign of $\chi$ is changed when analyzing $D^{-}$ decays to ensure $CP$ conservation. Neglecting the mass of the positron, the differential decay width for the $D^+ \to K_S^0\pi^0e^+\nu_e$ decay is written as~\cite{pwa-formalism}
\begin{equation}
\begin{aligned}
d^{5}\Gamma &~=~\frac{G^{2}_{F}||V_{cs}||^{2}}{(4\pi)^{6}m^{3}_{D}}X\beta{\cal I}(m^{2}, q^{2}, \theta_{K_{S}^0}, \theta_{e}, \chi)
\quad \times dm^{2} dq^{2} d\cos(\theta_{K_S^0}) d\cos(\theta_{e})d\chi.
\end{aligned}
\end{equation}
In this expression, $G_F$ is the Fermi constant, $|V_{cs}|$ is the $c \to s$ CKM matrix element, $X=p_{K_S^0\pi^{0}}m_{D^+}$, where $p_{K_S^0\pi^{0}}$ is the momentum of $K_S^0\pi^{0}$ in the $D^+$ rest frame, and $\beta= 2p^{*}/m$, where $p^{*}$ denotes the momentum of $K_S^0$ in the $K_S^0\pi^{0}$ rest frame.
The dependence of $\cal I$ on $\theta_{e}$ and $\chi$ is given by~\cite{pwa-formalism}
\begin{equation}
\begin{aligned}
  {\cal I}~=~& {\cal I}_{1}+{\cal I}_{2}\cos2\theta_{e}+{\cal I}_{3}\sin^{2}\theta_{e}\cos2\chi+{\cal I}_{4}
  \sin2\theta_{e}\cos\chi+\\
  &{\cal I}_{5}\sin\theta_{e}\cos\chi +{\cal I}_{6}\cos\theta_{e}+{\cal I}_{7}\sin\theta_{e}\sin\chi+
    {\cal I}_{8}\sin2\theta_{e}\sin\chi
  +{\cal I}_{9}\sin^{2}\theta_{e}\sin2\chi,
\end{aligned}
\label{eq:decay_intensity}
\end{equation}
where the quantities ${\cal I}_{1, \ldots,9}$ depend on $m^{2}$, $q^{2}$, and $\theta_{K_S^0}$. These quantities can be expressed in terms of three form factors ${\cal F}_{1,2,3}$:
\begin{equation}
  \begin{aligned}
  {\cal I}_{1}~=~&\frac{1}{4}\{|{\cal F}_{1}|^{2}+\frac{3}{2}\sin^{2}\theta_{K_S^0}
  (|{\cal F}_{2}|^{2}+|{\cal F}_{3}|^{2})\}, \\
  {\cal I}_{2}~=~&-\frac{1}{4}\{|{\cal F}_{1}|^{2}-\frac{1}{2}\sin^{2}\theta_{K_S^0}
  (|{\cal F}_{2}|^{2}+|{\cal F}_{3}|^{2})\}, \\
  {\cal I}_{3}~=~&-\frac{1}{4}\{|{\cal F}_{2}|^{2}-|{\cal F}_{3}|^{2}\}\sin^{2}\theta_{K_S^0}, \\
  {\cal I}_{4}~=~&\frac{1}{2}{\rm Re}({\cal F}^{*}_{1}{\cal F}_{2})\sin\theta_{K_S^0}, \\
  {\cal I}_{5}~=~&{\rm Re}({\cal F}^{*}_{1}{\cal F}_{3})\sin\theta_{K_S^0}, \\
  {\cal I}_{6}~=~&{\rm Re}({\cal F}^{*}_{2}{\cal F}_{3})\sin^{2}\theta_{K_S^0}, \\
  {\cal I}_{7}~=~&{\rm Im}({\cal F}_{1}{\cal F}_{2}^{*})\sin\theta_{K_S^0}, \\
  {\cal I}_{8}~=~&\frac{1}{2}{\rm Im}({\cal F}_{1}{\cal F}_{3}^{*})\sin\theta_{K_S^0}, \\
  {\cal I}_{9}~=~&-\frac{1}{2}{\rm Im}({\cal F}_{2}{\cal F}^{*}_{3})\sin^{2}\theta_{K_S^0}.
  \end{aligned}
\end{equation}
The form factors ${\cal F}_{i}$ can be expanded into partial waves
to show their explicit dependence on $\theta_{K_S^0}$. In this analysis, we only consider $S$ and $P$ waves, and thereby obtain
\begin{equation}
  \begin{aligned}
  {\cal F}_{1}&={\cal F}_{1S}+{\cal F}_{1P}\cos\theta_{K_S^0}, \\
  {\cal F}_{2}&=\frac{1}{\sqrt{2}}{\cal F}_{2P}, \\
  {\cal F}_{3}&=\frac{1}{\sqrt{2}}{\cal F}_{3P}. \\
  \label{eq:form_factor}
  \end{aligned}
\end{equation}
Here, ${\cal F}_{iS}$ and ${\cal F}_{iP}$ correspond to $S$ and $P$ waves.

\vspace{-0.0cm}

\subsection{$S$-WAVE FORM FACTOR}
\hspace{1.5em}
The $S$-wave related form-factor $F_{1S}$ is expressed as
\begin{equation}
 \begin{aligned}
  F_{10}=p_{K_S^0\pi^{0}}m_{D}\frac{1}{1-\frac{q^{2}}{m_{A}^{2}}}\mathcal{A}_{S}(m). \\
  \end{aligned}
\label{eq:form_factor_S}
\end{equation}
The $S$-wave amplitude $\mathcal{A}_{S}(m)$ is considered to be the superposition of  a non-resonant background and the $\bar K^{*}_{0}(1430)$ resonance. The phase parameterization used for the $S$-wave amplitude, as proposed by the LASS collaboration~\cite{Swavelass}, is given by
\begin{equation}
\cot(\delta_{BG}^{1/2})=\frac{1}{a_{S, BG}^{1/2}p^{*}}+\frac{b_{S, BG}^{1/2}p^{*}}{2},
\label{eq:flatte}
\end{equation}
\begin{equation}
\cot(\delta_{\bar K^{*}_{0}(1430)})=\frac{m^2_{\bar K^{*}_{0}(1430)}-m^2}{m_{\bar K^{*}_{0}(1430)}\Gamma_{\bar K^{*}_{0}(1430)}(m)},
\label{eq:flatte}
\end{equation}
\begin{equation}
\delta(m)=\delta_{BG}^{1/2} + \delta_{\bar K^{*}_{0}(1430)},
\label{eq:flatte}
\end{equation}
where $\delta_{BG}^{1/2} $ is the scattering length determined by the fit, $b_{S, BG}^{1/2}$ is the effective range fixed at $-0.81$~\cite{six1}, and $\delta_{\bar K^{*}_{0}(1430)}$ is the total phase of the $S$ wave.

The amplitude $\mathcal{A}_{S}(m)$ is described with formalism developed by the BaBar collaboration~\cite{five}. It is constructed as a piecewise function for $m$ below and above the $\mathcal{A}_{S}(m)$ pole-mass value:
\begin{equation}
\mathcal{A}_{S}(m)= r_{S}P(m)e^{i\delta(m)},m < m_{\bar K^{*}_{0}(1430)};
\end{equation}
\begin{equation}
\begin{aligned}
\mathcal{A}_{S}(m) = &r_{S}P(m_{\bar K^{*}_{0}(1430)})\times e^{i\delta(m)}\\
&\sqrt{\frac{(m_{\bar K^{*}_{0}(1430)}\Gamma_{\bar K^{*}_{0}(1430)})^2}{(m^2_{\bar K^{*}_{0}(1430)}-m^2)^2+(m_{\bar K^{*}_{0}(1430)}\Gamma_{\bar K^{*}_{0}(1430)})^2}},
 \quad m > m_{\bar K^{*}_{0}(1430)}.
\end{aligned}
\label{eq:flatte}
\end{equation}
In these expressions, $P(m)=1+r^{(1)}_S x$, $x = \sqrt{(\frac{m}{m_{K_S^0}+m_{\pi^0}})-1}$. $r_S$ is a dimensionless parameter, and the parameter $r_S^{(1)}$ represents  the relative strength of the $S$ wave, and $e^{i \delta(m)}$ introduces an overall phase to the $S$ wave. The magnitude is assumed to have a linear variation versus $m$ below the pole while behaving according to the Breit-Wigner function above the pole.

\subsection{$P$-WAVE FORM FACTOR}
\hspace{1.5em}
The $P$-wave related form-factors ${\cal F}_{iP}$ can be parameterized by the helicity form-factors $H_{0,\pm}$. These form factors are given by
\begin{equation}
  \begin{aligned}
  {\cal F}_{1P}=&2\sqrt{2}\alpha qH_{0}\times \mathcal{A}(m),  \\
  {\cal F}_{2P}=&2\alpha q(H_{+}+H_{-})\times \mathcal{A}(m), \\
  {\cal F}_{3P}=&2\alpha q(H_{+}-H_{-})\times \mathcal{A}(m). \\
  \label{eq:form_factor_P}
  \end{aligned}
\end{equation}
Here, the $\alpha$ value depends on the definition of the 
$P$-wave amplitude $\mathcal{A}(m)$, as shown in Eq.~(\ref{eq:alpha}). The helicity form-factors, related to two axial-vector
form-factors $A_{1,2}(q^{2})$ and one vector form-factor $V(q^{2})$, are given by
\begin{equation}
  \begin{aligned}
  H_{0}(q^{2})&=\frac{1}{2mq}[(m_{D}^{2}-m^{2}-q^{2})(m_{D}+m)A_{1}(q^{2})-4\frac{m_{D}^{2}p_{K_S^0\pi^{0}}^{2}}{m_{D}+m}A_{2}(q^{2})], \\
  H_{\pm}(q^{2})&=[(m_{D}+m)A_{1}(q^{2})\mp\frac{2m_{D}p_{K_S^0\pi^{0}}}{(m_{D}+m)}V(q^{2})].\\
  \label{eq:helicity}
  \end{aligned}
\end{equation}
The $q^{2}$ dependence is described using a single-pole dominance parametrization~\cite{SPD1,SPD2,SPD3}, which is written as
\begin{equation}
  \begin{aligned}
  V(q^{2})=\frac{V(0)}{1-\frac{q^{2}}{m_{V}^{2}}},\\
  A_{1}(q^{2})=\frac{A_{1}(0)}{1-\frac{q^{2}}{m_{A}^{2}}},\\
  A_{2}(q^{2})=\frac{A_{2}(0)}{1-\frac{q^{2}}{m_{A}^{2}}}.
  \end{aligned}
  \label{eq:spd}
\end{equation}
The pole masses $m_{V}$ and $m_{A}$ are
fixed at 1.81GeV/$c^{2}$ and 2.61 GeV/$c^{2}$~\cite{six1}, respectively.
The ratios of the form factors taken at $q^{2}=0$, $r_{V}=V(0)/A_{1}(0)$
and $r_{2}=A_{2}(0)/A_{1}(0)$, can reflect the variation of the differential decay rate versus the kinematic variables.
They are determined by the fit to the data.
The $P$-wave amplitude $\mathcal{A}(m)$ is defined as
\begin{equation}
\mathcal{A}(m)=\frac{m_{0}\Gamma_{0}F_{J}(m)}{m_{0}^{2}-m^{2}-im_{0}\Gamma(m)},
\label{eq:Am}
\end{equation}
where $m_0$ and $\Gamma_0$ are the pole mass and the total width of $\bar K^{*}(892)^{0}$, respectively. The width $\Gamma(m)$ is given by
\begin{eqnarray}
\Gamma(m)=\Gamma_{0}\frac{p^{*}}{p^{*}_{0}}\frac{m_{0}}{m}F_{J}^{2}(m), \\
F_{J}=\left(\frac{p^{*}}{p^{*}_{0}}\right)^{J}\frac{B_{J}(p^{*})}{B_{J}(p^{*}_{0})},
\label{eq:FL}
\end{eqnarray}
where $B_{J}$ is the Blatt-Weisskopf damping factor~\cite{rBW}.  In this case $J=1$ and  
\begin{equation}
B_{1} =1/\sqrt{1+r_{BW}^{2}p^{*2}}.
  \label{eq:b}
\end{equation}
Here, the $p^{*}$ is the momentum of the $K_S^0$ in the $K_S^0\pi^{0}$ rest frame, and
$p^{*}_{0}$ is the value taken at the pole mass of the resonance. A common value of $r_{BW}$ for light mesons is about 
0.5~fm, so it is fixed at 3.07 GeV$^{-1}$ in natural units~\cite{six1}.

The factor $\alpha$ entering in Eq.~(\ref{eq:form_factor_P}) is equal to
\begin{equation}
  \alpha = \sqrt{\frac{3\pi \mathcal{B}_{\bar K^{*}(892)^{0}}}{p^{*}_{0}\Gamma_{0}}},
  \label{eq:alpha}
\end{equation}
where $\mathcal{B}_{\bar K^{*}(892)^{0}}=\mathcal{B}(\bar K^{*}(892)^{0}\to \bar K^0\pi^0)$.

\subsection{AMPLITUDE-ANALYSIS METHOD}
\hspace{1.5em}
We utilize the unbinned maximum-likelihood method within the RooFit~\cite{RooFit-pwa} framework to perform the amplitude-analysis fit. The probability density function~(PDF) for a single candidate event is given by
\begin{equation}
  {\rm PDF}(\xi,\eta)= \frac{\omega(\xi,\eta)\epsilon(\xi)R_4(\xi)}{\int d\xi \omega(\xi,\eta) R_4(\xi)\epsilon(\xi)},
  \label{eq:pdf1}
\end{equation}
where $\xi$ is the final state of the event, and $\eta$ corresponds to the fit parameters in the PDF. Here, $\omega(\xi,\eta)$ denotes the decay intensity (as defined in Eq.~(\ref{eq:decay_intensity})), $\epsilon(\xi)$ is the reconstruction efficiency for the final state $\xi$, and $R_4(\xi)$ is the phase space factor.

The likelihood is expressed as the product of probabilities for all $N$ events. It is given by
\begin{equation}
  {\cal L}=\prod_{i=1}^{N}{\rm PDF}(\xi_{i}, \eta)
   =\prod_{i=1}^{N}\frac{\omega(\xi_i,\eta)\epsilon(\xi_i)R_4(\xi_i)}{\int d\xi_i \omega(\xi_i,\eta) R_4(\xi_i) \epsilon(\xi_i)}.
\end{equation}
\noindent

During the fitting process, the parameters $\eta$ are optimized by minimizing the negative log-likelihood~(NLL), which is written as
\begin{equation}
  -\ln\!{\cal L} =
  -\sum_{i=1}^{N}\ln(\epsilon(\xi_{i})R_4(\xi_i))
  -\sum_{i=1}^{N}\ln \frac{\omega(\xi_i,\eta)}{\int d\xi_i \omega(\xi_i,\eta) R_4(\xi_i)\epsilon(\xi_i)}.
  \label{eq:lnL}
\end{equation}
\noindent
The first term in the above equation depends only on the efficiency, and remains constant throughout the fit. Therefore, only the computation of the second term is performed in the fit. 
The NLL to be minimized can therefore be written as
\begin{equation}
  {\rm NLL}=-\sum_{i=1}^{N}\ln\frac{\omega(\xi_{i},\eta)}{\sigma},
  \label{eq:nll}
\end{equation}
\noindent
where  $\sigma = \int d\xi_i \omega(\xi_i, \eta) R_4(\xi_i)\epsilon(\xi_i)$.

The acceptance efficiency is taken into account in the calculation of the cross section $\sigma$, which is determined through MC integration with the signal MC sample~\cite{pwa-MCintegration}. Initially, we use phase space~(PHSP) MC samples with the uniform distribution of the kinematics to obtain the cross section $\sigma$. After obtaining the first set of solutions in the amplitude analysis, we update $\sigma$  from MC samples generated with this new set of solutions, and iterate until the solution stabilizes. 
The normalization integral term is given as
\begin{eqnarray}
\sigma=\int  \omega(\xi, \eta) \epsilon(\xi)R_4(\xi)d\xi
      \propto\frac{1}{N_{\rm selected}}\sum_{k=1}^{N_{\rm selected}} \frac{\omega~(\xi_{k}, \eta)}{\omega(\xi_{k}, \eta_{0})}.
\label{eq:sigmc-integral}
\end{eqnarray}
\noindent
Here, the terms $\eta$ and $\eta_{0}$ are the values of the parameters used in the fit
and that used to produce the simulated events, respectively.
$N_{\rm selected}$ denotes the number of the signal MC events after reconstruction and selection.

The background is estimated using the inclusive MC sample. It is subtracted from the total one via
\begin{eqnarray}
{\rm NLL}=  (-\ln L_{\rm total}-\omega(-\ln L_{\rm background})),
\label{neweq:sigmc-integral}
\end{eqnarray}
where $\omega$ is a normalization factor of the inclusive MC sample corresponding to the data size, with a value of 0.0185.

The goodness of the fit is estimated using $\chi^2/$n.d.f., where n.d.f. denotes the number of degrees of freedom. The $\chi^{2}$ is calculated by comparing the distribution in the five kinematic variables between data and the MC sample, determined by the results of the amplitude analysis. The $\chi^{2}$ value is determined by

\begin{equation}
\chi^2=\sum_i^{N_{\mathrm{bin}}}\frac{(n_i^{\mathrm{data}}-n_i^{\mathrm{fit}})^2}{n_i^{\mathrm{fit}}}.
\end{equation}

\begin{figure}[htbp]\centering
	\setlength{\abovecaptionskip}{-1pt}
	\setlength{\belowcaptionskip}{10pt}
\includegraphics[width=0.49\textwidth]{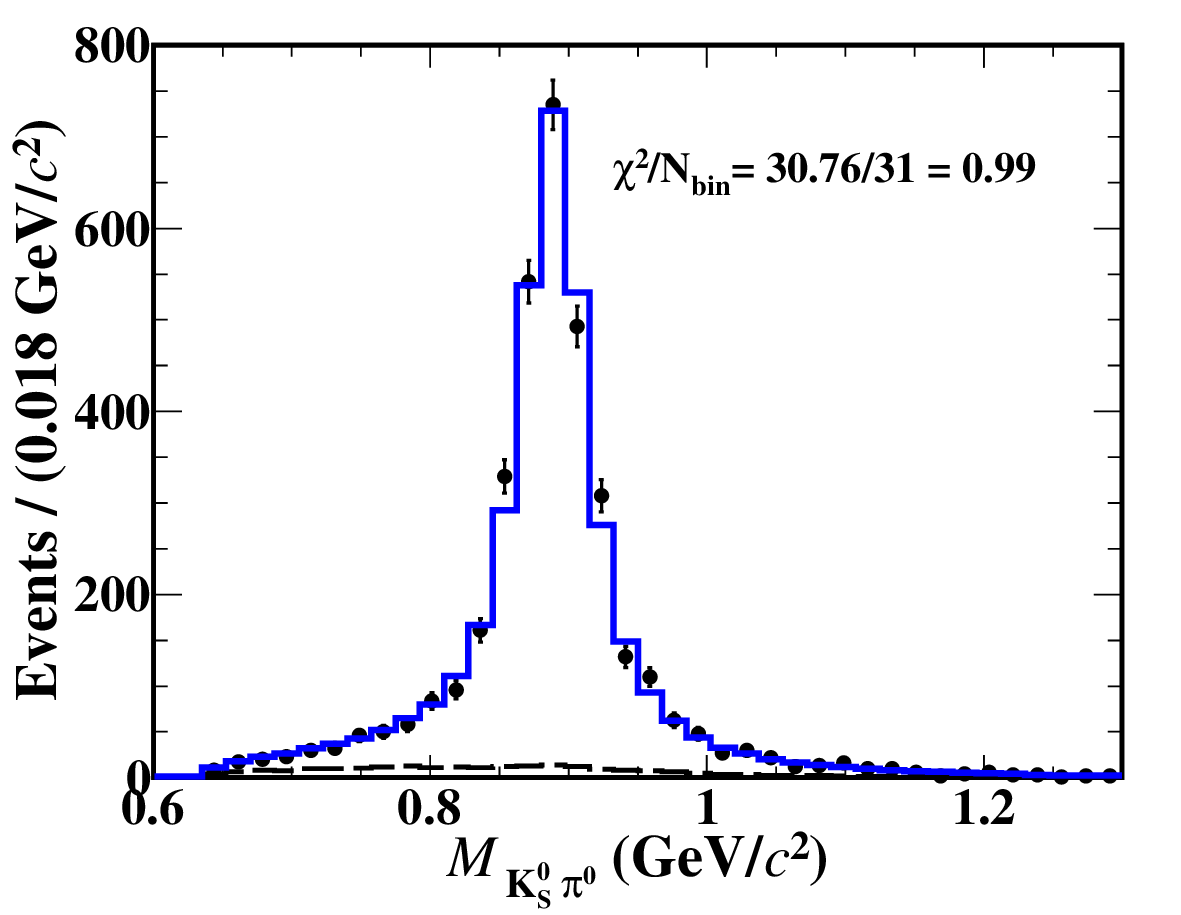}
\includegraphics[width=0.49\textwidth]{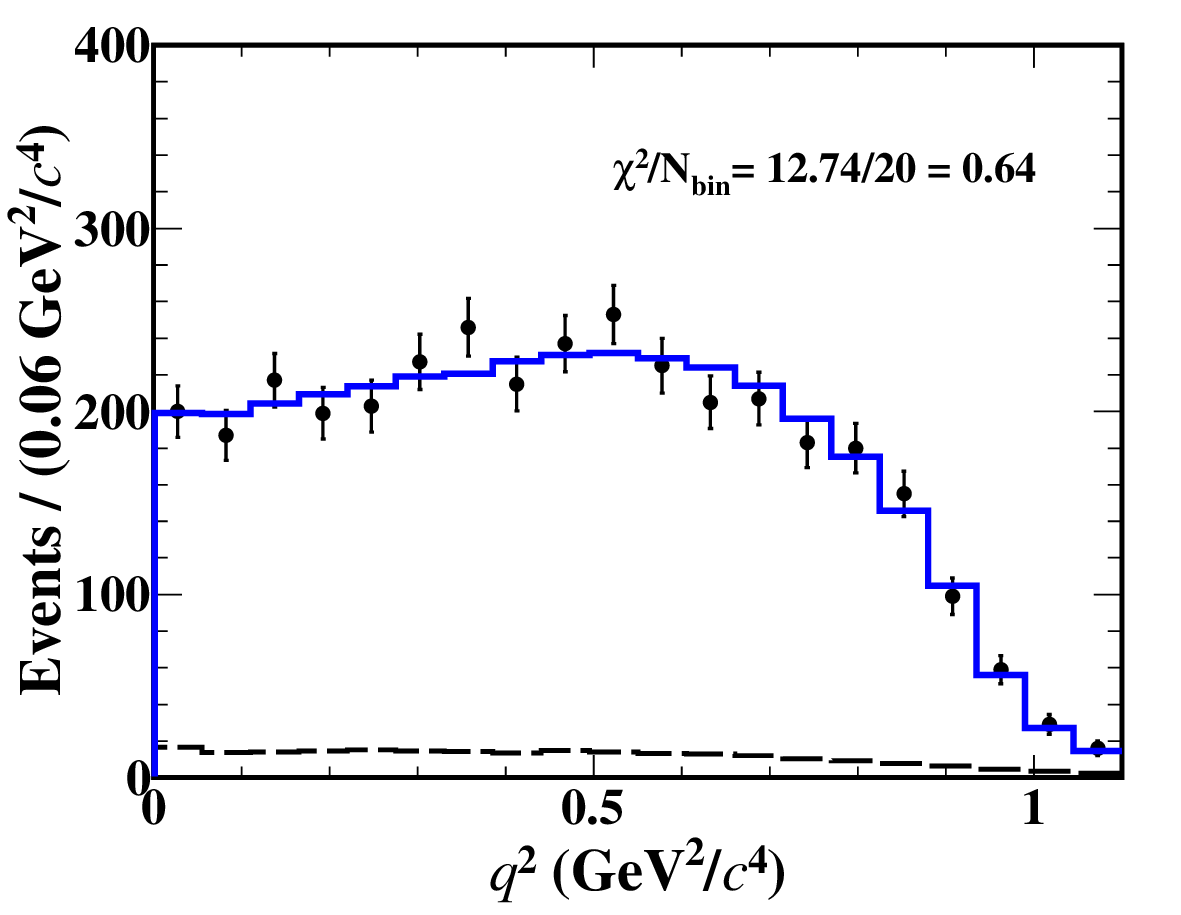} \\[0.1cm]
\includegraphics[width=0.49\textwidth]{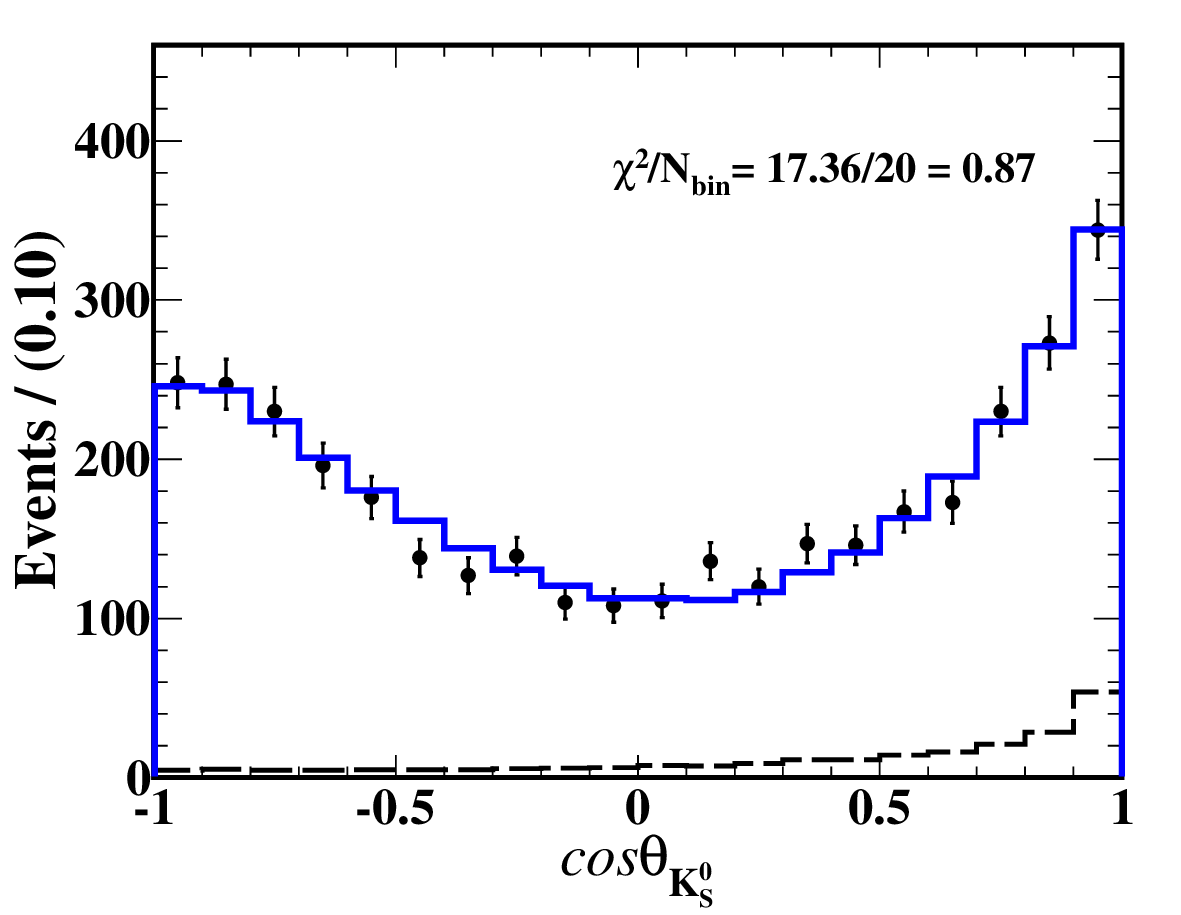}
\includegraphics[width=0.49\textwidth]{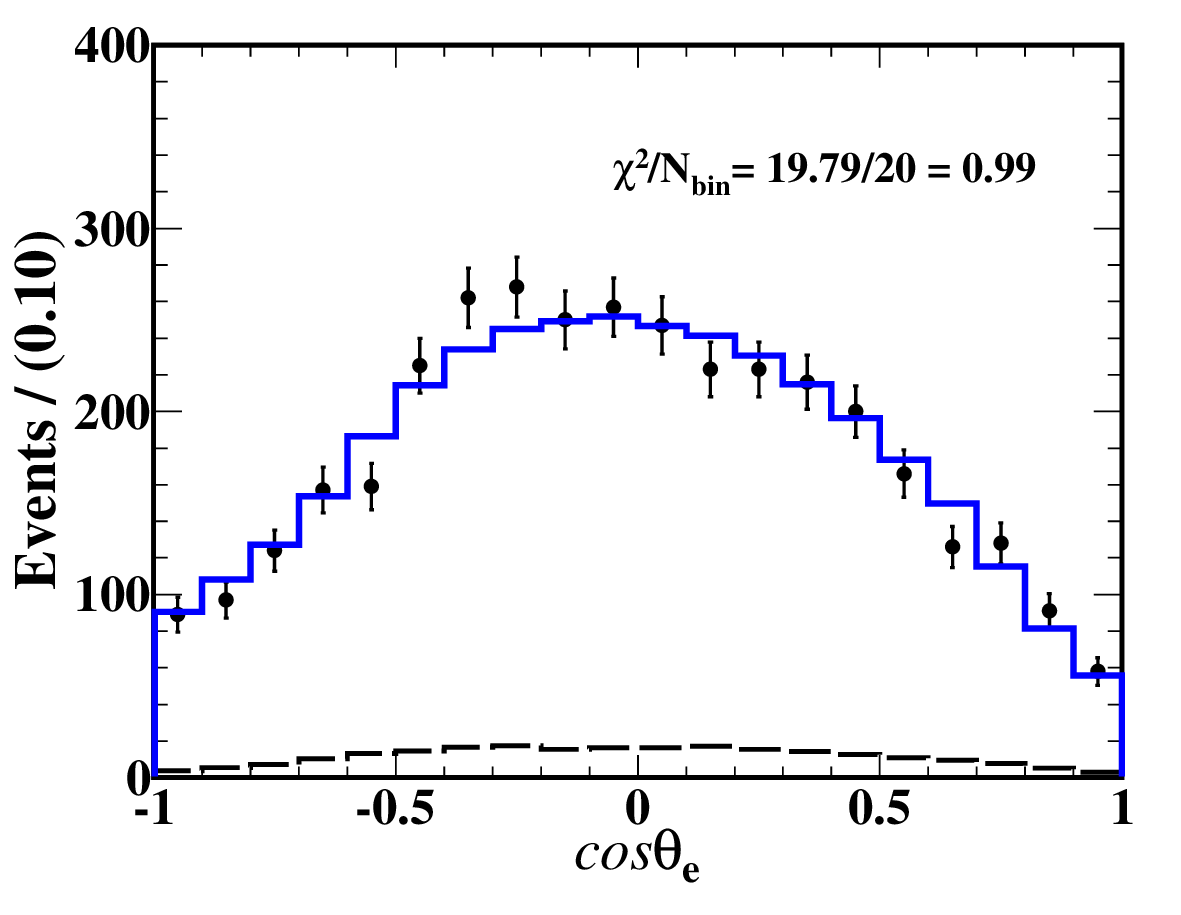} \\[0.1cm]
\includegraphics[width=0.49\textwidth]{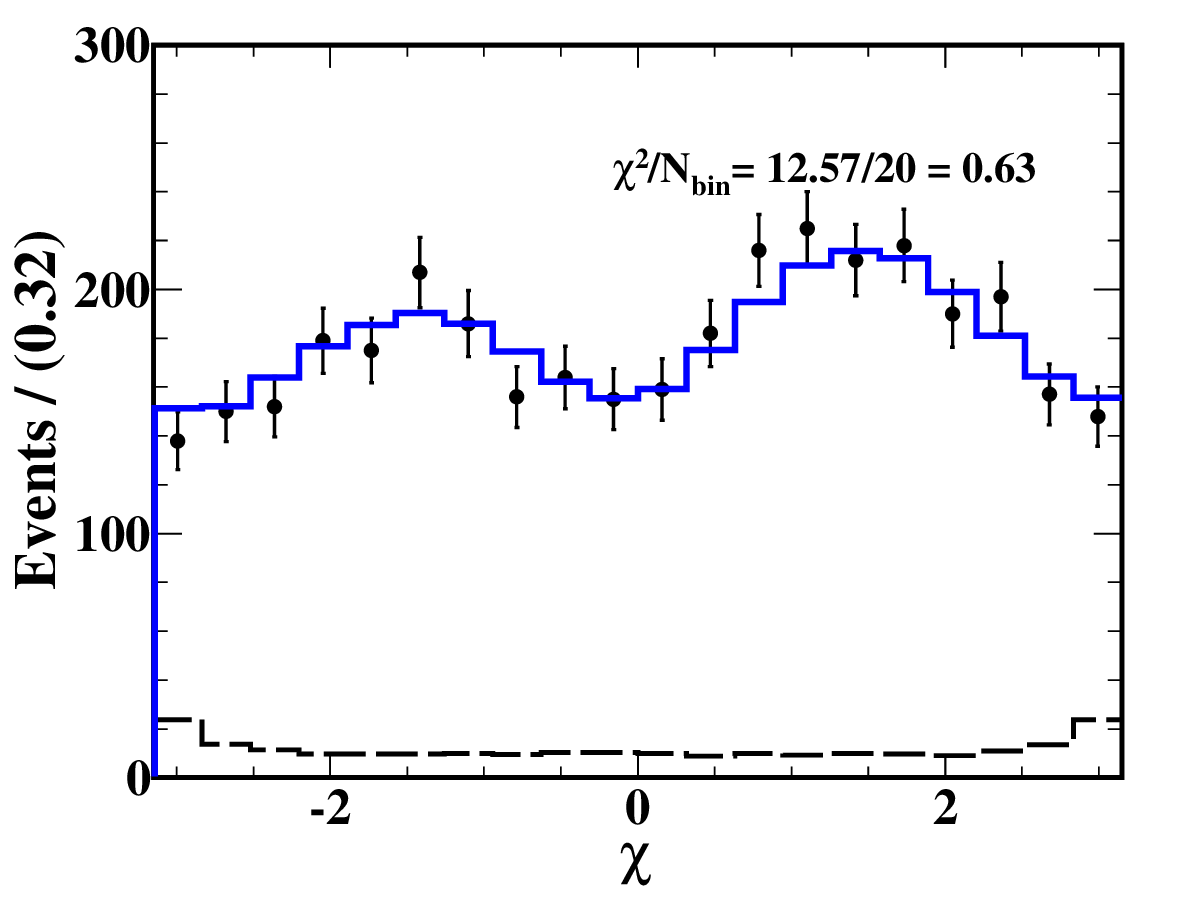}
\caption{Projections of the data and simultaneous amplitude analysis fit onto five kinematic variables for $D^+\to K_S^0\pi^0e^+\nu_e$.
The dots with error bars are data, the blue lines are the best fit, and the dashed lines show the sum of the simulated backgrounds.
For the $\chi^2/{\rm N_{bin}}$ calculation, we merge any neighboring bins with very few entries until there are 10 entries accumulated.
}
\label{fit:pwa-center}
\end{figure}

Based on the amplitude analysis solution, we calculate the fractions of each component according to
$f_{k}=\frac{\int d\xi \omega_{k}(\xi,\eta)}{\int d\xi \omega(\xi,\eta)}$,
which represents the ratio of the decay intensity of the specific component to the total intensity.
The $\omega_k(\xi,\eta)$ and $\omega(\xi,\eta)$ are the decay intensities of the $k$-th component and the total amplitude, respectively. In our analysis, we employ a large PHSP MC sample of $4\times 10^6$ events, without considering detector acceptance or resolution effects. To estimate the statistical uncertainty of the fraction, the fit results are iteratively adjusted based on the error covariance matrix of the fit, and new fractions are computed. Comparisons of the projections over the five kinematic variables between data and the amplitude analysis solution are illustrated in Fig.~\ref{fit:pwa-center}.
The fit results are summarized in Table~\ref{tab:ff-sum}.

\begin{table}[htbp]
\begin{center}
\caption{ The fitted parameters of the amplitude model under the assumption that the signal is composed of $S\text{-}{\rm wave}$ and $\bar K^{*}(892)^0$ components. The first and second uncertainties are statistical and systematic, respectively. }
\begin{tabular}{l|c}
\hline
\hline
Variable                                                     & Value   \\
\hline
$r_{2}$&$0.72\pm0.06\pm0.02$\\
$r_{V}$&$1.43\pm0.07\pm0.03$\\
$a^{1/2}_{S,BG }~(\rm GeV/\textit{c})^{-1}$&$2.04\pm0.17\pm0.17$\\
$m_{\bar K^{*}(892)^0}~(\rm {MeV}/\textit{c}^{2})$&$893.82\pm0.57\pm0.11$\\
$r_{S}~(\rm GeV)^{-1} $&$-8.34\pm0.24\pm0.30$\\
$\Gamma^0_{\bar K^{*}(892)^0}~(\rm {MeV}/\textit{c}^{2})$&$46.46\pm1.15\pm0.70$\\
$f_{S\text{-}{\rm wave}}$&$6.13\pm0.27\pm0.30$\\
$f_{\bar K^{*}(892)^0}$&$93.88\pm0.27\pm0.29$\\
\hline
\hline
\end{tabular}
\label{tab:ff-sum}
\end{center}
\end{table}

\vspace{-0.0cm}
\vspace{-0.0cm}

\subsection{SYSTEMATIC UNCERTAINTY OF AMPLITUDE ANALYSIS}
\hspace{1.5em}
The systematic uncertainties in the amplitude analysis are discussed below.
\begin{itemize}
\item[\bf (a)] {Background estimation}: The uncertainties associated with the  background contribution are studied by varying the fraction of the background events within its statistical uncertainty. The differences in fit result for each parameter are assigned as the corresponding systematic uncertainties.

\item[\bf (b)] {$r^{(1)}_{S}$, $b^{1/2}_{S,BG}$, and $r_{BW}$:} The uncertainties from the fixed parameters are estimated by varying the input values by $\pm 1\sigma$~\cite{six1} and re-performing the fit. The largest difference is taken as the systematic uncertainty due to the imperfect knowledge of each parameter.

\item[\bf (c)] {$m_V$ and $m_A$:}
The systematic uncertainties associated with the pole-mass assumption are estimated by varying $m_V$ and $m_A$ by $\pm 50~\text{MeV}/c^2$.~\cite{six1}.
The resulting differences compared to the baseline-fit solution are taken as the corresponding systematic uncertainties.

\item[\bf (d)] {Fit bias:} The possibility of  bias in the fit procedure is assessed by inspecting pull distributions from 
fits to 300 MC samples with a size equal to the data, where these samples considered detector acceptance and resolution effects. The means of the pulls are compatible with zero in all cases, but the widths are greater than unity by $10-20\%$ for certain parameters.   The statistical uncertainties of these parameters are scaled to reflect these slightly wider than expected pulls.

 \end{itemize}

Assuming that all sources are independent, the total systematic uncertainties on the amplitude analysis are determined by adding all uncertainties in quadrature.  The component systematic uncertainties and the totals are summarized in Table \ref{tab:sys_ff}.
\begin{table*}[htp]
\centering
\caption{The systematic uncertainties on the fitted parameters in the amplitude analysis~(in \%).}
\scalebox{0.85}{
\begin{tabular}{ccccccccc}
\hline
\hline
Source & $\Delta r_2$  &  $\Delta r_V$  &  $\Delta a^{1/2}_{S,BG}$ &$\Delta m_{\bar {K}^{*}{(892)}^0}$&$\Delta r_S$&$\Delta \Gamma^0_{\bar {K}^{*}{(892)}^0}$& $\Delta f_{\bar {K}^{*}{(892)}^0}$ & $\Delta f_{S{\text -}{\rm wave}}$   \\\hline
Background estimation&0.8 & 0.1 & 0.4 & 0.002 & 0.4 & 1.1 & 0.1 & 1.1\\\hline
$r_S^{(1)}$&0.7 & 1.1 & 0.4 & 0.005 & 3.1 & 0.4 & 0.2 & 3.4\\\hline
$b_{S,BG}^{1/2}$&1.3 & 0.2 & 8.0 & 0.004 & 0.5 & 0.4 & 0.1 & 0.8\\\hline
$r_{BW}$&1.4 & 0.6 & 1.0 & 0.008 & 1.8 & 0.8 & 0.2 & 3.1\\\hline
$m_{V}$&1.1 & 1.6 & 0.2 & 0.005 & 0.1 & 0.2 & 0.0 & 0.4\\\hline
$m_{A}$&1.5 & 0.5 & 0.4 & 0.004 & 0.2 & 0.2 & 0.0 & 0.2\\\hline
Fit bias&$-$ &$-$ &$-$&$-$&$-$&$-$&$-$&$-$\\\hline
Total& 2.8&2.0&8.1&0.012&3.6&1.5&0.3&4.8\\

\hline
\hline
\end{tabular}
\label{tab:sys_ff}
}
\end{table*}

\section{SUMMARY}
\begin{figure*}[htbp]\centering
	\setlength{\abovecaptionskip}{-1pt}
	\setlength{\belowcaptionskip}{10pt}
\includegraphics[width=7.4cm]{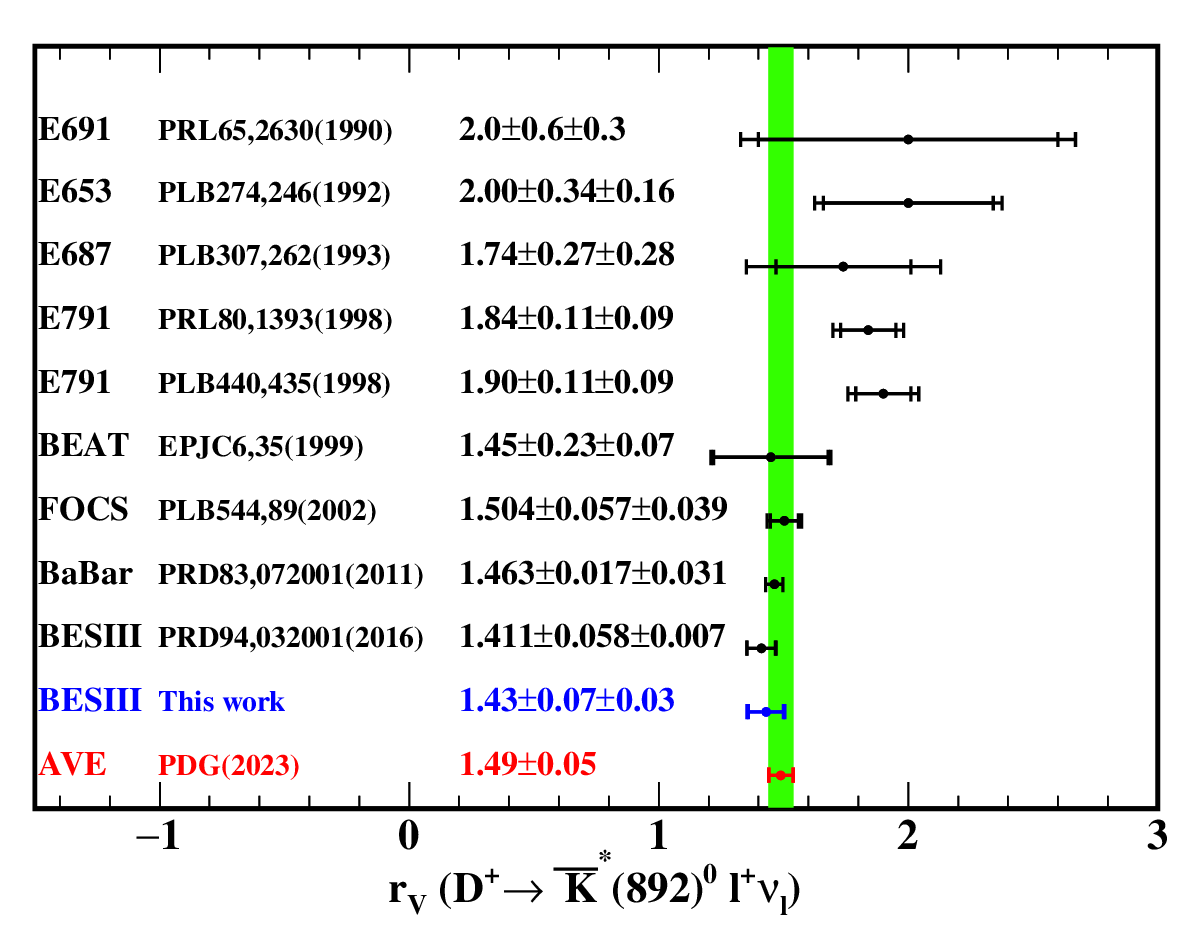}
\includegraphics[width=7.4cm]{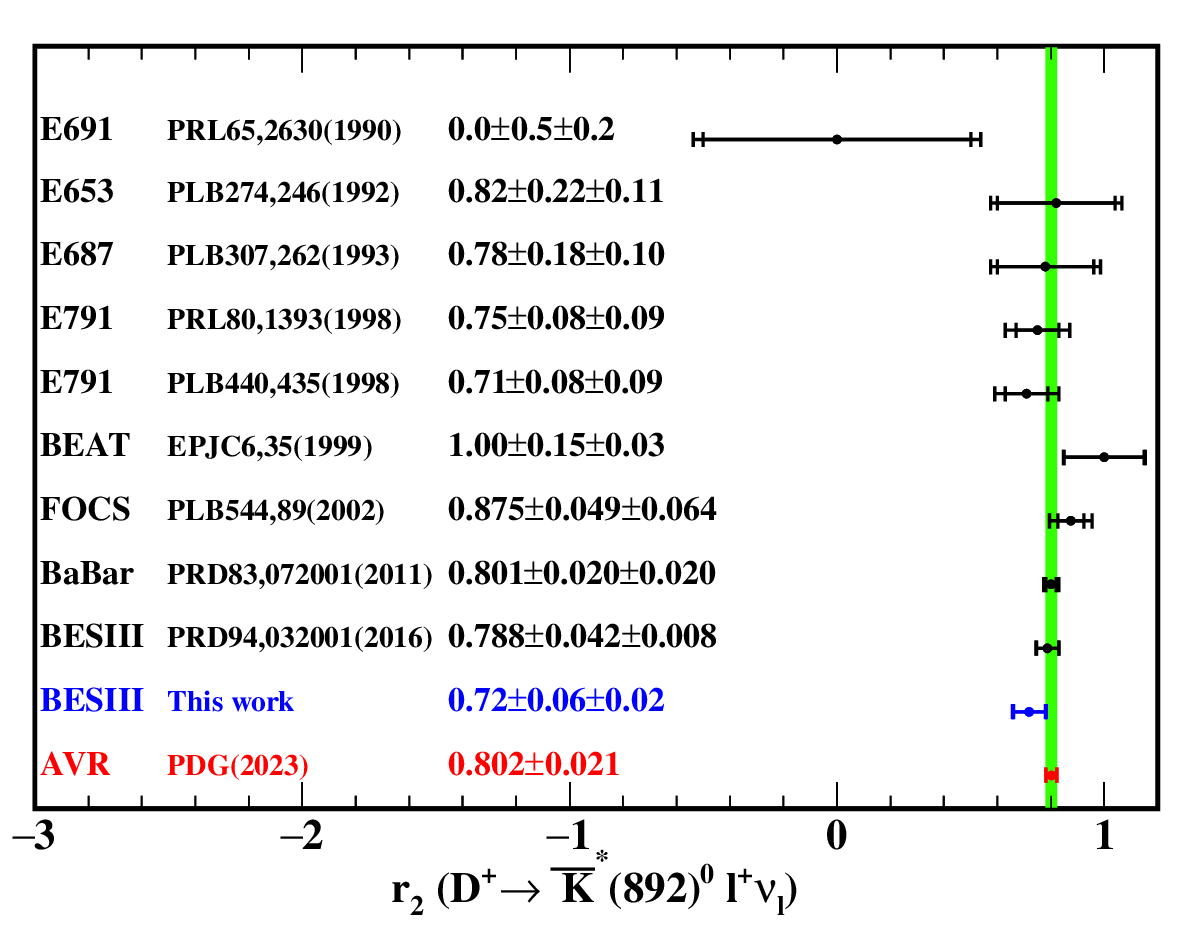}
\caption{Comparison of $r_V$ and $r_2$  with other measurements. The first and second uncertainties are statistical and systematic, respectively. The green band represents the average value in the PDG2023.}
\label{fig:pwa_comparison}
\end{figure*}

\begin{figure*}[htbp]\centering
	\setlength{\abovecaptionskip}{-1pt}
	\setlength{\belowcaptionskip}{10pt}
\includegraphics[width=11cm]{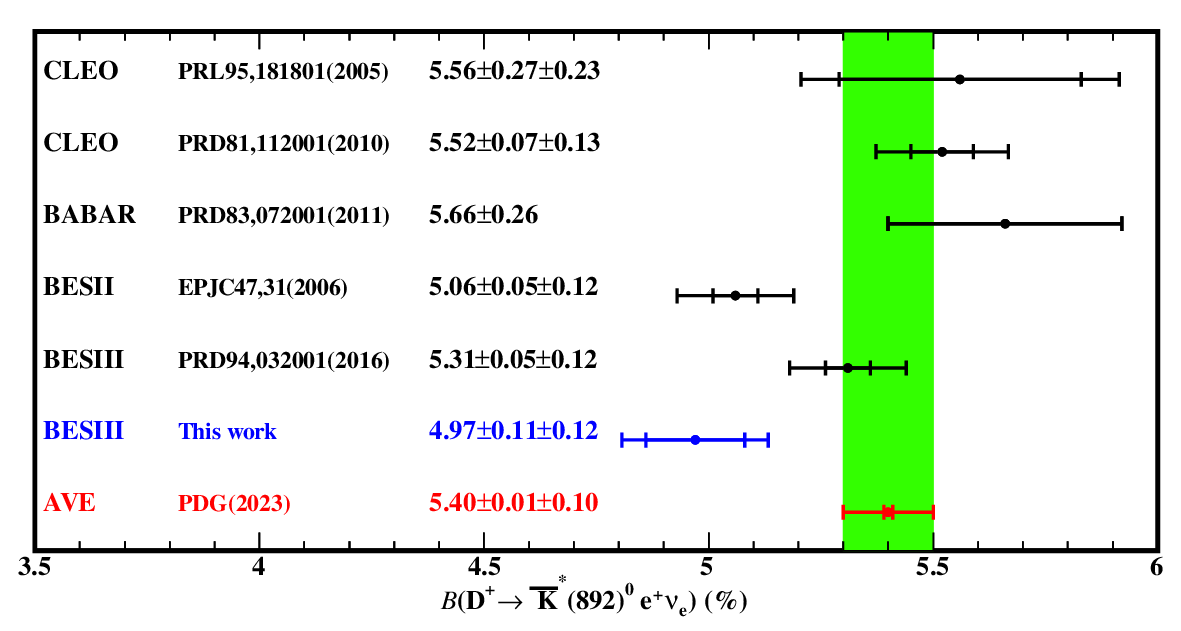}
\caption{Comparison of our BF of $D^+ \to \bar K^{*}(892)^{0}e^+\nu_e$ with other measurements. The first and second uncertainties are statistical and systematic, respectively. The green band represents the average value in the PDG2023.
}
\label{fig:kstarenu}
\end{figure*}

\hspace{1.5em}
Based on 7.93 $\rm fb^{-1}$ of $e^+e^-$ collision data taken at $\sqrt{s}=$ 3.773 GeV, the BF of $D^+\to K_{S}^{0} \pi^{0}e^+\nu_e$ is measured for the first time and determined to be ${\mathcal B}$($D^+\to K_S^0\pi^0e^+\nu_e$) =
$(0.881\pm0.017_{\rm stat.}\pm0.016_{\rm syst.})$\%. From the analysis of the $D^+\to K_S^0\pi^0e^+\nu_e$ decay dynamics, the $S\text{-}{\rm wave}$ and $P$-wave components are determined with fractions of $f_{S\text{-}{\rm wave}}$ = $(6.13 \pm 0.27_{\rm stat.} \pm 0.30_{\rm syst.})\%$, and $f_{\bar K^{*}(892)^0}$ = $(93.88 \pm 0.27_{\rm stat.} \pm 0.29_{\rm syst.})$\%, respectively. From these results the BFs of the two components are calculated to be ${\mathcal B}$($D^+\to (K_S^0\pi^0)_{S\text{-}{\rm wave}}~e^+\nu_e$) = $(5.41 \pm 0.35_{\rm stat.} \pm 0.37_{\rm syst.})\times10^{-4}$ and ${\mathcal B}$($D^+\to \bar K^{*}(892)^0e^+\nu_e$) = $(4.97 \pm 0.11_{\rm stat.} \pm 0.12_{\rm syst.})$\%, where ${\mathcal B}(\bar K^*(892)^0 \to \bar K^0 \pi^0) \times {\mathcal B}(\bar K^0 \to K_S^0 )=1/6$ based on isospin conservation with negligible uncertainty.
Additionally, the hadronic form-factor ratios of the $D^+\to \bar K^{*}(892)^0e^+\nu_e$ decay are determined to be $r_V$ = $1.43 \pm 0.07_{\rm stat.} \pm 0.03_{\rm syst.}$ and $r_2$ = $0.72 \pm 0.06_{\rm stat.} \pm  0.02_{\rm syst.}$. Figure~\ref{fig:pwa_comparison} shows the comparisons of $r_V$ and $r_2$  with different experimental results. Figure~\ref{fig:kstarenu} shows the comparison of our BF of $D^+ \to \bar K^{*0}(892)e^+\nu_e$ with other measurements. Our results are consistent with those reported in other experiments~\cite{MARK-III:1990bbt,four, five, six1}. Table~\ref{tab:theo} shows the hadronic form factors of $D\to \bar K^*$ predicted by different theoretical calculations, as well as the values measured in the current study. The measured results of $r_V$ and $r_2$ are consistent with several Lattice QCD calculations within 2$\sigma$~\cite{Lubicz:1992,Bernard:1991bz,Abada:1994,Bowler:1994zr,APE:1994kxx}, but disfavor those  that employ a large lattice spacing~\cite{Bhattacharya:1994db}, as well as those predictions based on the  quark model~\cite{Wirbel:1985ji,Isgur:1988gb,Gilman:1989uy} and QCD sum rules~\cite{theo:Sum}.

\begin{table}[htp]
\centering
\caption{Hadronic form factors of $D\to \bar K^*$ at $q^2$=0 predicted by different theories.}
\scalebox{0.9}{
\begin{tabular}{|c|c|c|c|}
\hline
\multicolumn{2}{|c|}{Reference} & $r_V$ & $r_2$ \\\hline
Experiment                    & Average~\cite{ref::pdg2022} &1.49$\pm$0.05&0.802$\pm$0.021\\ \hline
\multirow{5}{*}{Lattice QCD}& LMMS~\cite{Lubicz:1992}  &1.6$\pm$0.2            &0.4$\pm$0.4\\
                              & BKS~\cite{Bernard:1991bz}&1.99$\pm$0.22$\pm$0.33 &0.7$\pm$0.16$\pm$0.17\\
                              & ELC~\cite{Abada:1994}  &1.3$\pm$0.2              &0.6$\pm$0.3\\
                              & UKQCD~\cite{Bowler:1994zr}&$1.4^{+0.5}_{-0.2}$   &0.9$\pm$0.2\\
                              & LANL~\cite{Bhattacharya:1994db} &1.83$\pm$0.09   &0.74$\pm$0.19\\
                              & APE~\cite{APE:1994kxx}&1.6$\pm$0.3&0.7$\pm$0.4\\ \hline
\multirow{3}{*}{Quark model}  & WSB~\cite{Wirbel:1985ji}  &1.4&1.3\\
                              & ISGW~\cite{Isgur:1988gb}  &1.4&1.0\\
                              & GS~\cite{Gilman:1989uy}   &2.0&0.8 \\ \hline
QCD sum rules                 & BBD~\cite{theo:Sum}       &2.2$\pm$0.2&1.2$\pm$0.2\\ \hline
This work                     & $-$     &$1.43 \pm 0.07_{\rm stat.} \pm 0.03_{\rm syst.}$&$0.72 \pm 0.06_{\rm stat.} \pm  0.02_{\rm syst.}$\\ \hline
\end{tabular}
}
\label{tab:theo}
\end{table}

\vspace{-0.0cm}
\vspace{-0.0cm}

\vspace{-0.0cm}

\acknowledgments
\hspace{1.5em}
The BESIII Collaboration thanks the staff of BEPCII and the IHEP computing center for their strong support. This work is supported in part by National Key R\&D Program of China under Contracts Nos. 2023YFA1606000, 2020YFA0406300, 2020YFA0406400; National Natural Science Foundation of China (NSFC) under Contracts Nos. 11635010, 11735014, 11935015, 11935016, 11935018, 12025502, 12035009, 12035013, 12061131003, 12192260, 12192261, 12192262, 12192263, 12192264, 12192265, 12221005, 12225509, 12235017, 12361141819; Natural Science Foundation of Hunan Province (Contracts No.2024JJ2044); Guangdong Basic and Applied Basic Research Foundation under Grant No. 2023A1515010121 and the Fundamental Research Funds for the Central Universities under Contract No. 020400/531118010467; the Chinese Academy of Sciences (CAS) Large-Scale Scientific Facility Program; the CAS Center for Excellence in Particle Physics (CCEPP); Joint Large-Scale Scientific Facility Funds of the NSFC and CAS under Contract No. U1832207; CAS Key Research Program of Frontier Sciences under Contracts Nos. QYZDJ-SSW-SLH003, QYZDJ-SSW-SLH040; 100 Talents Program of CAS; The Institute of Nuclear and Particle Physics (INPAC) and Shanghai Key Laboratory for Particle Physics and Cosmology; European Union's Horizon 2020 research and innovation programme under Marie Sklodowska-Curie grant agreement under Contract No. 894790; German Research Foundation DFG under Contracts Nos. 455635585, Collaborative Research Center CRC 1044, FOR5327, GRK 2149; Istituto Nazionale di Fisica Nucleare, Italy; Ministry of Development of Turkey under Contract No. DPT2006K-120470; National Research Foundation of Korea under Contract No. NRF-2022R1A2C1092335; National Science and Technology Fund of Mongolia; National Science Research and Innovation Fund (NSRF) via the Program Management Unit for Human Resources \& Institutional Development, Research and Innovation of Thailand under Contract No. B16F640076; Polish National Science Centre under Contract No. 2019/35/O/ST2/02907; The Swedish Research Council; U. S. Department of Energy under Contract No. DE-FG02-05ER41374.

\bibliographystyle{IEEEtran}

\newpage
\begin{small}
\begin{center}
M.~Ablikim$^{1}$, M.~N.~Achasov$^{4,c}$, P.~Adlarson$^{76}$, O.~Afedulidis$^{3}$, X.~C.~Ai$^{81}$, R.~Aliberti$^{35}$, A.~Amoroso$^{75A,75C}$, Q.~An$^{72,58,a}$, Y.~Bai$^{57}$, O.~Bakina$^{36}$
% ...
\\
\vspace{0.2cm}
(BESIII Collaboration)\\
\vspace{0.2cm} {\it
$^{1}$ Institute of High Energy Physics, Beijing 100049, People's Republic of China\\
% ...
$^{3}$ Bochum  Ruhr-University, D-44780 Bochum, Germany\\
$^{4}$ Budker Institute of Nuclear Physics SB RAS (BINP), Novosibirsk 630090, Russia\\
% ...
$^{35}$ Johannes Gutenberg University of Mainz, Johann-Joachim-Becher-Weg 45, D-55099 Mainz, Germany\\
$^{36}$ Joint Institute for Nuclear Research, 141980 Dubna, Moscow region, Russia\\
% ...
$^{57}$ Southeast University, Nanjing 211100, People's Republic of China\\
$^{58}$ State Key Laboratory of Particle Detection and Electronics, Beijing 100049, Hefei 230026, People's Republic of China\\
% ...
$^{72}$ University of Science and Technology of China, Hefei 230026, People's Republic of China\\
% ...
$^{75}$ University of Turin and INFN, (A)University of Turin, I-10125, Turin, Italy; (B)University of Eastern Piedmont, I-15121, Alessandria, Italy; (C)INFN, I-10125, Turin, Italy\\
$^{76}$ Uppsala University, Box 516, SE-75120 Uppsala, Sweden\\
% ...
$^{81}$ Zhengzhou University, Zhengzhou 450001, People's Republic of China\\
\vspace{0.2cm}
$^{a}$ Deceased\\
% ...
$^{c}$ Also at the Novosibirsk State University, Novosibirsk, 630090, Russia\\
% ...
}\end{center}
\end{small}

\end{document}